\documentclass[%
 amsmath,amssymb,
 twocolumn,
 aps,
pre,
floatfix
]{revtex4-2}

\usepackage{graphicx}
\usepackage{dcolumn}
\usepackage{mathtools}

\usepackage{bm}
\usepackage{xcolor}
\usepackage{hyperref}
\usepackage{systeme}
\usepackage{multirow}
\usepackage{makecell}
\usepackage[normalem]{ulem}

\begin{document}

\preprint{APS/123-QED}

\title{The Learning by Confusion approach to identification of discontinuous phase transitions}%

\author{Monika Richter--Laskowska}
\affiliation{%
 University  of  Silesia,  Katowice,  Poland
}%
 \affiliation{Łukasiewicz Research Network – Krakow Institute of Technology, The Centre for
Biomedical Engineering, Zakopiańska 73, Kraków, 30-418, Poland}
 \email{monika.richter-laskowska@kit.lukasiewicz.gov.pl}

\author{Marcin Kurpas}%
\affiliation{%
 Institute of Physics, University of Silesia in Katowice, 41-500 Chorzów, Poland
}%

\author{Maciej M. Maśka}

\affiliation{
Institute of Theoretical Physics, Wrocław University of Science and Technology, Wrocław, Poland
}%

\date{\today}

\begin{abstract}

  Recently, the {\it Learning by Confusion} (LbC) approach has been proposed as a machine learning tool to determine the critical temperature $T_c$ of phase transitions without any prior knowledge of its even approximate value. The method has been proven effective, but it has been used only for continuous phase transitions, where the {\it confusion} results only from deliberate incorrect labeling of the data. However, in the case of a discontinuous phase transition, additional {\it confusion} can result from the coexistence of different phases. To verify whether the confusion scheme can also be used for discontinuous phase transitions, we apply the LbC method to three microscopic models, the Blume-Capel, the $q$--state Potts, and the Falicov-Kimball models, which undergo continuous or discontinuous phase transitions depending on model parameters. With the help of a simple model, we predict that the phase coexistence present in discontinuous phase transitions can indeed make the neural network more {\it confused} and thus decrease its performance. However, numerical calculations performed for the models mentioned above indicate that other aspects of this kind of phase transition are more important and can render the LbC method even less effective. Nevertheless, we demonstrate that in some cases the same aspects allow us to use the LbC method to identify the order of a phase transition. 
  
\end{abstract}

\maketitle


\section{\label{sec:intro}Introduction}

Over the last decade, the growth of computational power and the development of new algorithms have helped machine learning (ML) methods to gain great popularity in various domains.
They are widely applied to image recognition, natural language processing, or medical diagnostics \cite{Jordan255}. The ability to identify, classify, and interpret unusual patterns also makes them suitable for solving condensed matter physics problems, which, due to the exponentially large Hilbert space, are computationally expensive \cite{review}. 

Machine learning methods perform especially well in acceleration of Monte Carlo simulations \cite{slmc-1, slmc-2, slmc-3, slmc-4, wang-mc, Wu2019, Sharir2020}, they can provide an alternative representation of complex quantum wave functions \cite{Carleo, Glasser2018}, and are capable of identifying classical~\cite{melko, wang, wetzel, crossovers, Shiina2020, Ponte2017, Giannetti2018}, topological~\cite{topological-1, topological-2, topological-3, topological-4, topology2} and quantum phase transitions~\cite{quantum-1, quantum-2, quantum-3, khatami}.
The last application, that is, finding the critical temperature ($T_c$), is usually based on supervised learning, where a neural network is trained using a set of labeled configurations (e.g, spin configurations). The labels indicate whether a given configuration represents the low- or high-temperature phase. However, the problem is that to assign labels, one has to know at least approximately the value of the critical temperature.

To overcome this difficulty, a scheme called {\it learning by confusion} (LbC) has been proposed \cite{confusion}.
This method allows one to determine the transition temperature without any prior knowledge of its even approximate value. As such, it can be classified as an unsupervised learning method, but it exploits supervised techniques with data that are deliberately labeled incorrectly. The idea is based on the observation that the performance of a neural network is best if the data are labeled correctly and decreases with increasing discrepancy between the true and assumed critical temperatures. 
The applicability of the LbC method has been demonstrated in the study of classical spin models \cite{confusion}, frustrated magnetic models \cite{confusion5}, nuclear gas--liquid \cite{ Wang2020}, systems with double phase transitions and quasi-long-range order \cite{confusion2},
complex networks \cite{confusion3}, transitions in polariton lattices \cite{polaritons}, many-body localization in quantum chains \cite{PhysRevB.98.174202, lbc}, the entanglement breakdown of quantum states \cite{entanglement}, and ferrimagnetic alloys \cite{Koritsky2020}.

Modern classification of phase transitions distinguishes two types of transitions, namely discontinuous (first--order) and continuous (second--order) transitions~\cite{stanley1972introduction, dierking2003physics}.
The distinction between these two is quite obvious in the thermodynamic limit: some physical quantities, such as heat capacity or susceptibility, become divergent at the critical point in the case of first--order phase transition. However, this phenomenon is not observable when it comes to the second--order phase transition, for which the transition is always smooth.

In numerical simulations, the discrimination between continuous and discontinuous transitions can be difficult because all singularities in the temperature dependence of physical quantities disappear due to the finite size of a system. Instead, finite peaks occur at a critical point in both types of transitions~\cite{barber}.
To overcome this issue, one can look at the energy or magnetization histograms close to the transition temperature or calculate the Binder--Challa--Landau cumulant~\cite{binder_challa, landau,challa, bindercz, shumpei}. However, these methods require finite-size scaling, which can be very computationally demanding.
A good example here is the Falicov--Kimball (FK) model ~\cite{fk_model} tackled with the Monte Carlo approach. Since the Hamiltonian of the FK model contains coupled quantum and classical degrees of freedom, it has to be diagonalized in each Monte Carlo iteration. As a result, the limits for the maximum size of systems available for numerical simulations are lower, and finite-size scaling is more difficult. The application of this technique is necessary to properly establish the character of a phase transition, for example, using the histogram and the Binder--Challa--Landau cumulants methods. 
The latter, however, does not perform equally well for all microscopic models. For instance, it has been demonstrated in Ref.~\cite{Oliveira2019, supp} that for the FK model the cumulant method cannot be used to determine the critical interaction $U^{*}$ separating the continuous and discontinuous phase transitions.

A promising alternative to the approaches mentioned above is the use of machine learning methods, which can offer a substantial increase in computational performance compared to classical methods \cite{potts-neural,crossovers}. Thus, in this paper, we focus on one of the machine learning algorithms, namely the LbC scheme, and verify its applicability to the problem of classification of phase transitions. 
Based on the thermodynamic properties of phase transitions close to $T_c$ we introduce an intuitive theoretical model that describes the performance of a neural network for first- and second-order phase transitions. 
Next, using the LbC method, we perform a numerical analysis of phase transitions for the FK model, the Blume-Capel (BC) model \cite{bc1, bc2}, and the q-state Potts (\textit{q}P) model \cite{potts_model} and juxtapose the results with the predictions of the theoretical model.
We show that for BC and \textit{q}P models the neural network correctly identifies the critical temperature independently of the type of phase transition. For the FK model, $T_c$ is correctly determined only in the case of a continuous phase transition. For a discontinuous phase transition, the prediction accuracy shows a large plateau at a value close to 1, which leads to the ambiguity of $T_c$. The existence of this plateau cannot be explained by the characteristic for discontinuous phase transitions coexistence of low- and high-temperature phases, as we show with the help of the theoretical model. 
Our results indicate that the LbC scheme finds aspects of discontinuous phase transitions which are weakly connected to characteristic thermodynamic quantities.

The paper is organized as follows. In Section \ref{sec:TheIdea} we introduce a theoretical model for neural network prediction accuracy. Then, in Section \ref{sec:models}, we describe microscopic model Hamiltonians to which we apply the LbC scheme. In Section \ref{sec:cumulants} we discuss the standard methods of phase transition characterization. Section \ref{sec:results} explains the details of the calculation methods and the results obtained. Section \ref{sec:summary} contains a discussion of the results and the final conclusions.

\section{The learning by confusion method and phase coexistence}
\label{sec:TheIdea}
The LbC algorithm was specifically designed to recognize phase transitions in physical models \cite{confusion}. It is a hybrid algorithm that combines the features of supervised and unsupervised learning. In this scheme, a neural network goes through a training phase, but the training dataset is not always correctly labeled, leading to the \textit{confusion} of the neural network.
In the following, we summarize the key steps of the algorithm. A detailed discussion can be found in Ref. \cite{confusion}.

\begin{figure}
    \centering
    \includegraphics[width=0.9 \columnwidth]{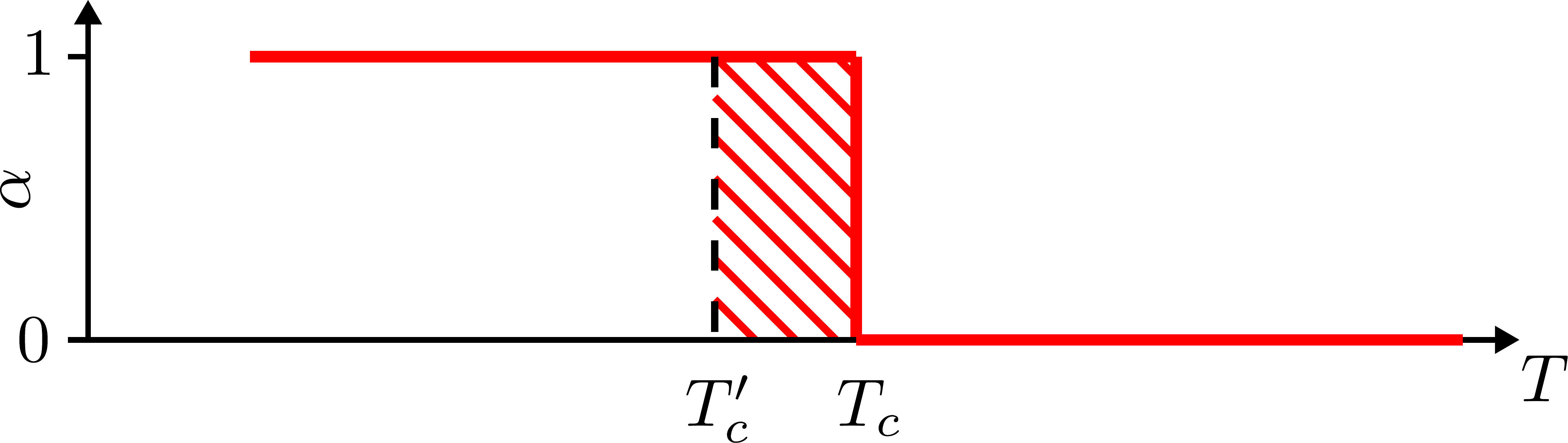}
    \caption{The fraction of the low-temperature phase $\alpha$ as a function of temperature in the case of a continuous phase transition at $T_c$. $T_c'$ indicates the fictitious critical temperature used to label the data. The hatched area, showing the amount of incorrectly labeled data, increases linearly with increasing distance between $T_c$ and $T_c'$.}
    \label{fig:continuous}
\end{figure}

Suppose that the input dataset includes samples generated in the temperature range $\left[T_{\min}, T_{\max}\right]$ and that the critical temperature $T_c$ lies between $T_{\min}$ and $T_{\max}$. In the learning process, a fictitious critical temperature $T'_{c}$ is assumed and the samples are labeled as it was the true critical temperature. Depending on how far $T'_{c}$ is from the real $T_{c}$, different amounts of data will be incorrectly labeled, as marked in Fig.~\ref{fig:continuous} by the hatched rectangular.

In the next step, the data are used to train the neural network and the performance of the resulting model $P(T'_{c})$ is determined. The whole training process is repeated for several different values of $T'_{c}$ lying in the range from $T_{\min}$ to $T_{\max}$ and $P(T'_{c})$ is plotted. For many models, $P(T'_{c})$ has a characteristic W--shape and the true $T_c$ can then be found from the position of the central maximum of $P(T'_{c})$. 

There are many different metrics used during the evaluation of ML classification models (accuracy, precision, recall, etc.). In the simple analysis presented in the current section, we will measure the performance of the neural network using the simplest metrics, i.e. {\it accuracy}, defined as the number of correct predictions divided by the total number of samples. Later, when analyzing numerical results, we will use the {\it area under receiver operating characteristic curve} (AUC--ROC) measure. 

\subsection{Continuous phase transitions}
We start our analysis with the case for which the LbC scheme has been proposed, that is, with a continuous phase transition. For this kind of transition, the two distinct phases do not coexist. This situation is depicted in Fig.~\ref{fig:continuous}.  
As demonstrated there, the amount of incorrectly labeled data increases {\it linearly} with the distance $|T_c-T_c'|$, and thus one can expect that the performance $P(T'_{c})$  {\it linearly} decreases when moving away from $T_c$. Indeed, it has been shown in Ref.~\cite{confusion} that for some models $P(T'_{c}) \propto |T_c-T_c'|$. 

To make the discussion more concrete, let us introduce a measure of the amount of incorrectly labeled data as:
\begin{equation}
    {\cal A}_{T_c'}=\int^{T_{\max}}_{T_{\min}}\zeta(T_c',T)dT,
    \label{eq:calA}
\end{equation}
where $\zeta(T_c',T)$ is the fraction of data points at temperature $T$ that have incorrect labels for a given $T_c'$. Performance $P(T'_{c})$ decreases with an increase of ${\cal A}_{T_c'}$ and, for the sake of simplicity, we assume that for $T_c'\approx T_c$ $P(T'_{c})=1-{\cal A}_{T_c'}/\delta T$, where $\delta T\equiv T_{\rm max}-T_{\rm min}$. This assumption guarantees that if the neural network is trained with randomly labeled data, one obtains $P(T_c')=0.5$, i.e., the probability that it correctly classifies any configuration is $0.5$. 

\begin{figure}[h]
    \centering
    \includegraphics[width = \columnwidth]{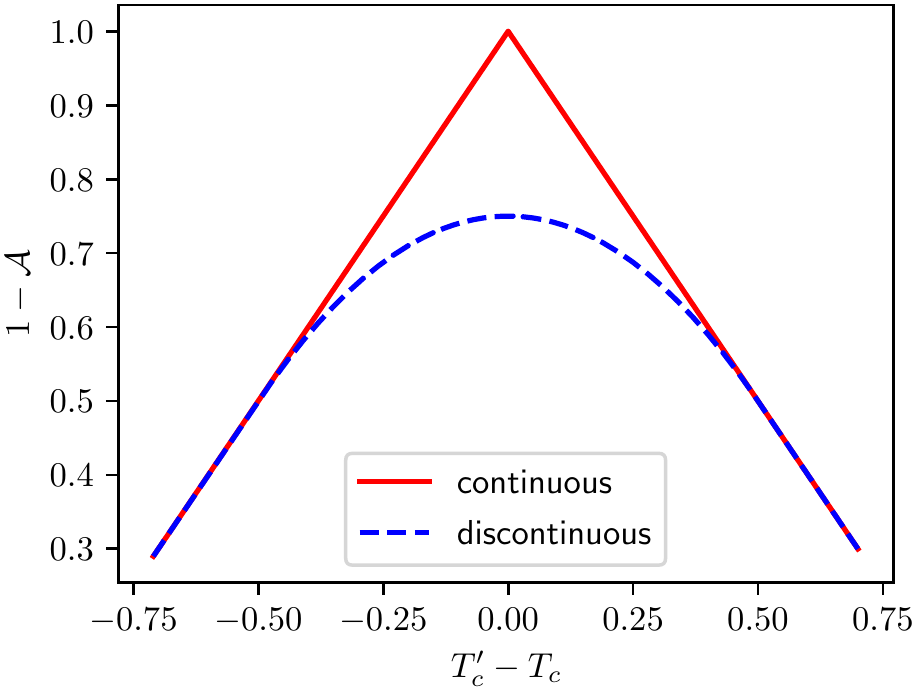}
    \caption{The performance $P(T_c')\equiv 1-{\cal A}_{T_c'}/\delta T$ obtained from the simple toy model introduced for continuous and discontinuous phase transitions. The amount of incorrectly labeled data is for a continuous transition ${\cal A}_{T_c'} \propto |T_c'-T_c|$, whereas for a discontinuous one it is given by Eq.~(\ref{eq:model_A}).}
    \label{fig:model_A}
\end{figure}

Let us denote by $\alpha$ the fraction of the system occupied by the low-temperature phase. For a continuous transition, there is no phase coexistence and therefore
\begin{equation}
    \alpha(T)=\begin{cases}
    1 & \mbox{for}\ \ T < T_c,\\
    0 & \mbox{otherwise}.
    \end{cases}
    \label{eq:simple_model}
\end{equation}
Then, as can be seen in Fig.~\ref{fig:continuous},
\begin{equation}
    \hspace*{-1mm}\zeta(T_c',T)=\begin{cases}
    1 & \mbox{for}
    \ \ T_c'<T<T_c\ \ \mbox{or}\ \ T_c<T<T_c',\\
    0 & \mbox{otherwise}.
    \end{cases}
    \label{eq:simple_model1}
\end{equation}
By inserting Eq.~(\ref{eq:simple_model1}) into Eq.~(\ref{eq:calA}), one obtains ${\cal A}_{T_c'}$ which increases linearly with the distance between $T_c'$ and $T_c$. It is shown by the red line presented in Fig.~\ref{fig:model_A}. This is true only in the thermodynamic limit, where the phase transition is sharp. However, for any finite system, a finite width of the phase transition will decrease the accuracy. It can be seen in works where the LbC method has been applied \cite{confusion, Koritsky2020, confusion2, confusion3, confusion5}.

\subsection{Discontinuous phase transitions}
The situation is different for a discontinuous phase transition, where the low- and high-temperature phases coexist around the critical point. To demonstrate that the neural network is more ``confused'' in this case than in the case of a continuous phase transition, we propose a simple model. We do not present or develop a theory of discontinuous phase transitions but propose a toy model that can be a rough approximation of the situation in numerical simulations for finite-size systems. It is independent of the phase conversion mechanism, i.e., whether it is spinodal decomposition or nucleation, and its validity will later be verified by Monte Carlo simulations for three different model Hamiltonians.

Let us assume that below $T_1$ only the low-temperature phase exists in the system. Then, when the temperature exceeds $T_1$, a fraction of the high-temperature phase is formed and increases linearly with the temperature up to $T_2$, where the high-temperature phase occupies the entire system. Denoting by $\alpha$ the fraction of the low-temperature phase ($1-\alpha$ is the fraction of the high-temperature phase) one can write:
\begin{equation}
    \alpha(T)=\begin{cases}
    1 & \mbox{for}\ \ T\le T_1,\\[2pt]
    \dfrac{T_2-T}{T_2-T_1} & \mbox{for}\ \ T_1 <T\le T_2,\\[7pt]
    0 & \mbox{for}\ \ T_2\le T.
    \end{cases}
    \label{eq:simple_model2}
\end{equation}
Between $T_1$ and $T_2$ both phases coexist and this is the temperature range in which the energy distribution has two maxima, one of them increasing and the other decreasing as the temperature changes from $T_1$ to $T_2$. The equal magnitude of these peaks signals $T_c$. In our model, we assume $T_c=\frac{1}{2}(T_1+T_2)$. Note that the model can be applied only to finite--size systems since in the thermodynamic limit, both $T_1$ and $T_2$ would converge to $T_c$.

\begin{figure}[h]
    \centering
    \includegraphics[width =0.9 \columnwidth]{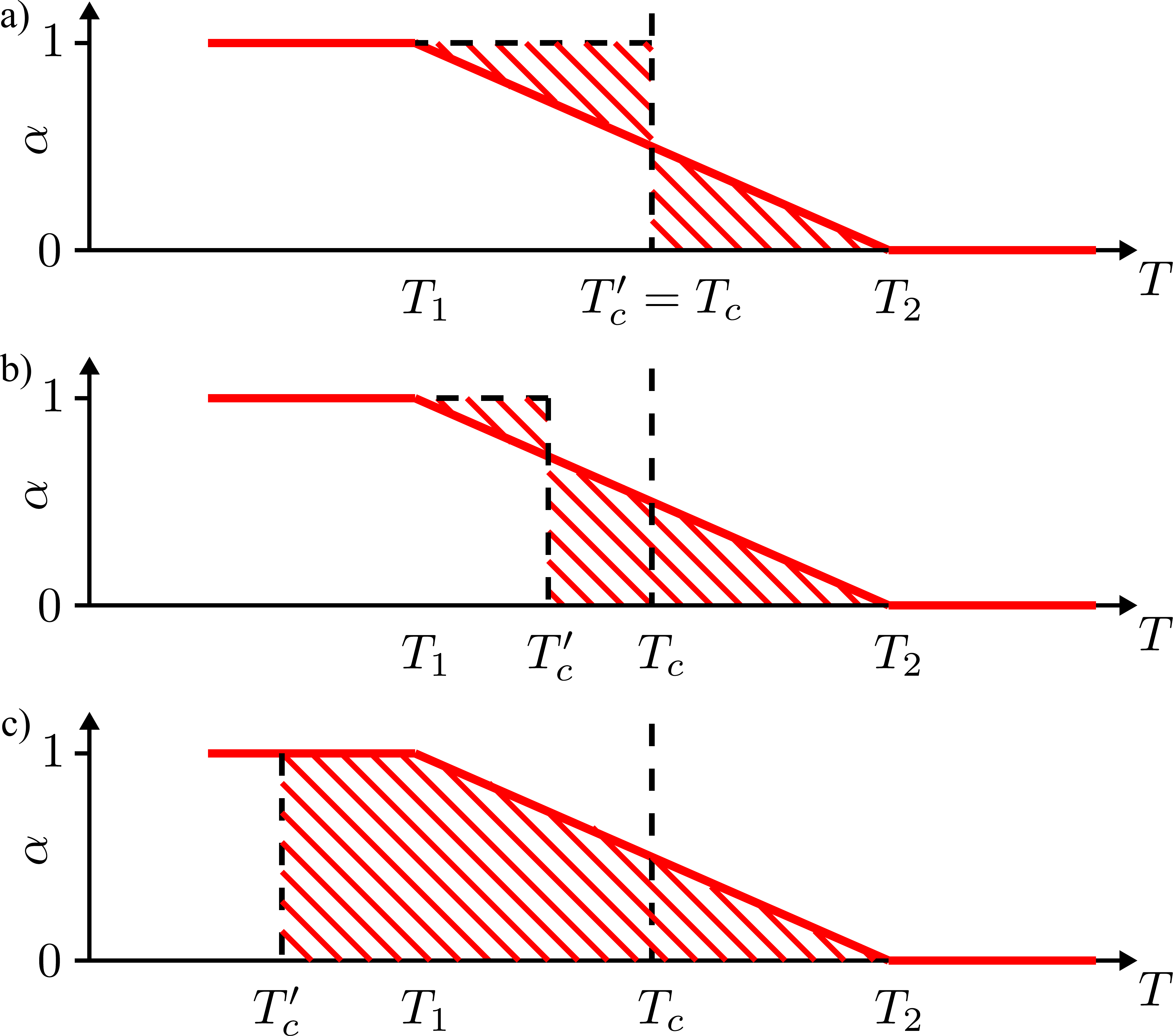}
    \caption{The same as in Fig.~\ref{fig:continuous}, but for a discontinuous phase transition. In the simple model, it is assumed that the low- and high-temperature phases coexist between $T_1$ and $T_2$. As in Fig.~\ref{fig:continuous}, the hatched area is proportional to the amount of incorrectly labeled data. The panels show situations corresponding to different values of the fictitious critical temperature: $T_c'=T_c$ (a), $T_1<T_c'<T_c$ (b), and $T_c'<T_1$ (c).}
    \label{fig:discontinuous}
\end{figure}

As can be seen in Fig.~\ref{fig:discontinuous}a, within the framework of this model, even for $T_c'=T_c$ there are some incorrectly labeled data points. Indeed, at $T_c$ there is a mixture of equal numbers of configurations from the low- and high-temperature phases. Therefore, independently of the attached labels, 50\% of the configurations will be incorrectly labeled.
When the distance between $T_c'$ and $T_c$ increases, ${\cal A}_{T_c'}$ also increases. The explicit form of $\zeta(T_c',T)$ depends on the relation between $T,\:T_c,\:T_c',\:T_1$, and $T_2$ (see Fig.~\ref{fig:discontinuous}) and will not be given here, but it can easily be inferred from Fig.~\ref{fig:discontinuous} that the integral shown in Eq.~(\ref{eq:calA}) gives
\begin{equation}
    {\cal A}_{T_c'}=\begin{cases}\dfrac{1}{\Delta}(T_c'-T_c)^2+\dfrac{\Delta}{4} & \mbox{ for }\ |T_c'-T_c|<\frac{1}{2}\Delta,\\[3mm]
    |T_c'-T_c| & \mbox{ for }\ |T_c'-T_c|>\frac{1}{2}\Delta,
    \end{cases}
    \label{eq:model_A}
\end{equation}
where $\Delta\equiv T_2-T_1$. This dependence is shown by the blue dashed line in Fig.~\ref{fig:model_A}. 

The intuitive toy model given by Eq.~(\ref{eq:simple_model}) has been introduced only to illustrate the idea and to estimate the difference in accuracy between continuous and discontinuous phase transitions. As can be seen there, one can expect a well-pronounced maximum around $T_c$ in the case of the continuous phase transition and a rather flat temperature dependence in the case of the discontinuous transition. 

In the following, we verify this idea for the FK, BC, and \textit{q}P models in which, depending on the model parameters, a continuous or discontinuous phase transition occurs. In particular, we demonstrate that the proposed explanation is too simple to precisely describe the shape of the accuracy vs. temperature dependence.  Nevertheless, numerical calculations show that there are cases where the LbC approach can still be used to recognize the order of the phase transition, although the underlying mechanism is different. It is important that this method is most efficient for the FK model, where it is not easy to determine the kind of phase transition. Therefore, most of the work will be related to this model.

\section{Models and Methods}\label{sec:models}

In order to verify the hypothesis described in the preceding section, we applied the LbC scheme to three models, which in different parameters' regimes exhibit either continuous or discontinuous phase transitions. Although most of the work focuses on the FK model, a similar analysis was also carried out for the \textit{q}P and BC models. Numerical calculations were performed for two-dimensional square lattices with periodic boundary conditions. Most of the results presented in the main text are obtained for $16\times 16$ systems. Nevertheless, we verify the results for lattices of different sizes, from $10\times 10$ to $26\times 26$. The comparison shows that the difference between the behavior for continuous and discontinuous transitions is only quantitatively affected by the system size. Moreover, as presented in Appendix \ref{sec:fss}, the main effect we present in this paper becomes more pronounced with increasing size.

\subsection{The Falicov--Kimball model} 
The Falicov--Kimball model was originally proposed to explain the metal-semiconductor transition in SmB$_6$ and metal oxides \cite{fk_model}. It describes the interaction of itinerant light particles (Bloch electrons) with heavy, localized ones. The latter can represent different physical objects, such as localized ions, dopants, localized ($f$) electrons or heavy atoms in optical lattices. In the following, we will use the term ``ion'' for the heavy particles. In the second quantization language, the Hamiltonian of the FK model can be written as
\begin{equation}\label{fk_hamiltonian}
\mathcal{H} = -t\sum_{\langle i,j\rangle}c_{i}^{\dag}c_{j} + U\sum_{i}n_{i}w_{i}.
\end{equation}
The first term in Eq.~\eqref{fk_hamiltonian} describes the kinetic energy of itinerant (spinless) electrons with $t$ denoting the nearest-neighbor hopping energy and $c^{\dagger}_i$ ($c_{i}$) representing the creation (annihilation) of an electron at lattice site $i$. The second term is responsible for the on--site electron-ion Coulomb interaction with $U$ and $n_{i}$ denoting the Coulomb potential and the electron particle number operator, respectively. $w=0$ ($w=1$) indicates that the site $i$ is unoccupied (occupied) by an ion. 
Although the Hamiltonian (\ref{fk_hamiltonian})  does not include a direct coupling between ions, there exists an effective long-range interaction mediated by electrons. 
This effective inter-ionic interaction leads to the self-organization of ions which, in turn,  gives rise to temperature--driven order-disorder phase transition. Finite-temperature Monte Carlo simulations have shown that for sufficiently small values of the Coulomb potential $U/t$ the phase transition is discontinuous, while for interactions stronger than a critical value $U^{*}/t \approx 1$ the transition changes its character and becomes continuous \cite{fk_maska}.

\subsection{The $q$--state Potts model}\label{subsec:Potts_model}
The $q$--state Potts model (\textit{q}P model) is  a generalization of the Ising model for $q$ possible spin orientations (projections) \cite{potts_model} determined by the angle $\theta_{n}$,
\begin{equation}
\label{spin_orientations}
    \theta_{n} = \frac{2\pi n}{q}, \quad n=0,1,\dots, q-1.
\end{equation}
Two spins interact only if they point in the same direction on nearest neighboring sites,
\begin{equation}
    \mathcal{H} = -J\sum_{\langle i,j\rangle}\delta_{s_{i}, s_{j}},
\end{equation}
where $s_{i(j)}=0, \dots, q-1$ represents the spin direction at site $i$ ($j$) given by $\theta_{i}$ ($\theta_{j}$), $\delta$ is the Kronecker delta, and the summation runs over all pairs of nearest neighbors. 
The coupling between spins leads to an order-disorder phase transition, which is of the first order for $q>4$ and second order for $q\le 4$ \cite{Baxter_1973, potts_renormalization}.

\subsection{The Blume--Capel model}
The third model we discuss here is the Blume--Capel (BC) model. It can be viewed as the Ising model for spin $S=1$, in which the external magnetic field that polarizes the spins along the $z$ axis is replaced by an anisotropy field $D$ \cite{bc1, bc2},
\begin{equation}
    \mathcal{H} = -J\sum_{\langle i,j\rangle}s_{i}s_{j} + D\sum_{i}s_{i}^2.
    \label{H_BC}
\end{equation}
The first term in Eq. (\ref{H_BC}) denotes the usual spin--spin interaction between neighboring sites, with $J$ denoting the exchange energy, while the second term describes the interaction of spins with the anisotropy field $D$.

The system undergoes a continuous or discontinuous phase transition, depending on the value of $D$ \cite{bc-phase, bc-phase2}. The phase diagram of the BC model contains a tricritical point, which, according to the predictions obtained from Monte Carlo simulations, is estimated in the thermodynamic limit as $k_{B}T_{c}/J=0.609(4),\; D_{c}/J=1.965(5)$~\cite{bc-mc}.

\subsection{Neural networks and datasets}\label{sec:datasets}

To demonstrate that the LbC method can be used to determine the kind of phase transition, we apply it to the three models described above. From a broad parameter space, we choose two representative sets for which the systems undergo continuous and discontinuous phase transitions. 
These parameters are listed in Table~\ref{tab:table1}.
\begin{table}
    \centering
    \setlength{\tabcolsep}{10pt}
    \renewcommand{\arraystretch}{2}
    \begin{tabular}{|c|>{\centering}m{2cm}|>{\centering\arraybackslash}m{2cm}|}
    \hline
    \multirow{2}{2em}{model} & \multicolumn{2}{c|}{parameters}\\
    & \makecell{continuous \\ transition} & \makecell{discontinuous \\ transition} \\[2mm]
    \hline\hline
    Falicov-Kimball (FK) & $U=4$ & $U=0.5$\\
    \hline
    $q$-state Potts (\textit{q}P) & $q=2$ & $q=10$ \\
    \hline
    Blume-Capel (BC) & $D=0$ & $D=1.98$ \\
    \hline
    \end{tabular}
    \caption{Model parameters used in this study.}
    \label{tab:table1}
\end{table}
The energy units in our calculations are set by the hopping integral $t=1$ for the FK model and by $J=1$ for the \textit{q}P  and BC models.
The training data for the neural network are generated with classical MC simulations for the \textit{q}P and BC models and with a combination of MC and exact diagonalization for the FK model. For each temperature, we generate $10^4$ statistically independent configurations. 
The data are sent to a convolutional neural network with a single convolutional layer. More details of the architecture of the neural network are included in Appendix \ref{app:details}.
When training the neural network, we use the Area Under the ROC Curve (AUC--ROC) as the metrics of model performance $P(T'_{c})$ \cite{ROC}.

\section{Determination of the nature of  the phase transitions}\label{sec:cumulants}

\begin{figure} [ht]
    \centering
    \includegraphics[width=0.99\columnwidth]{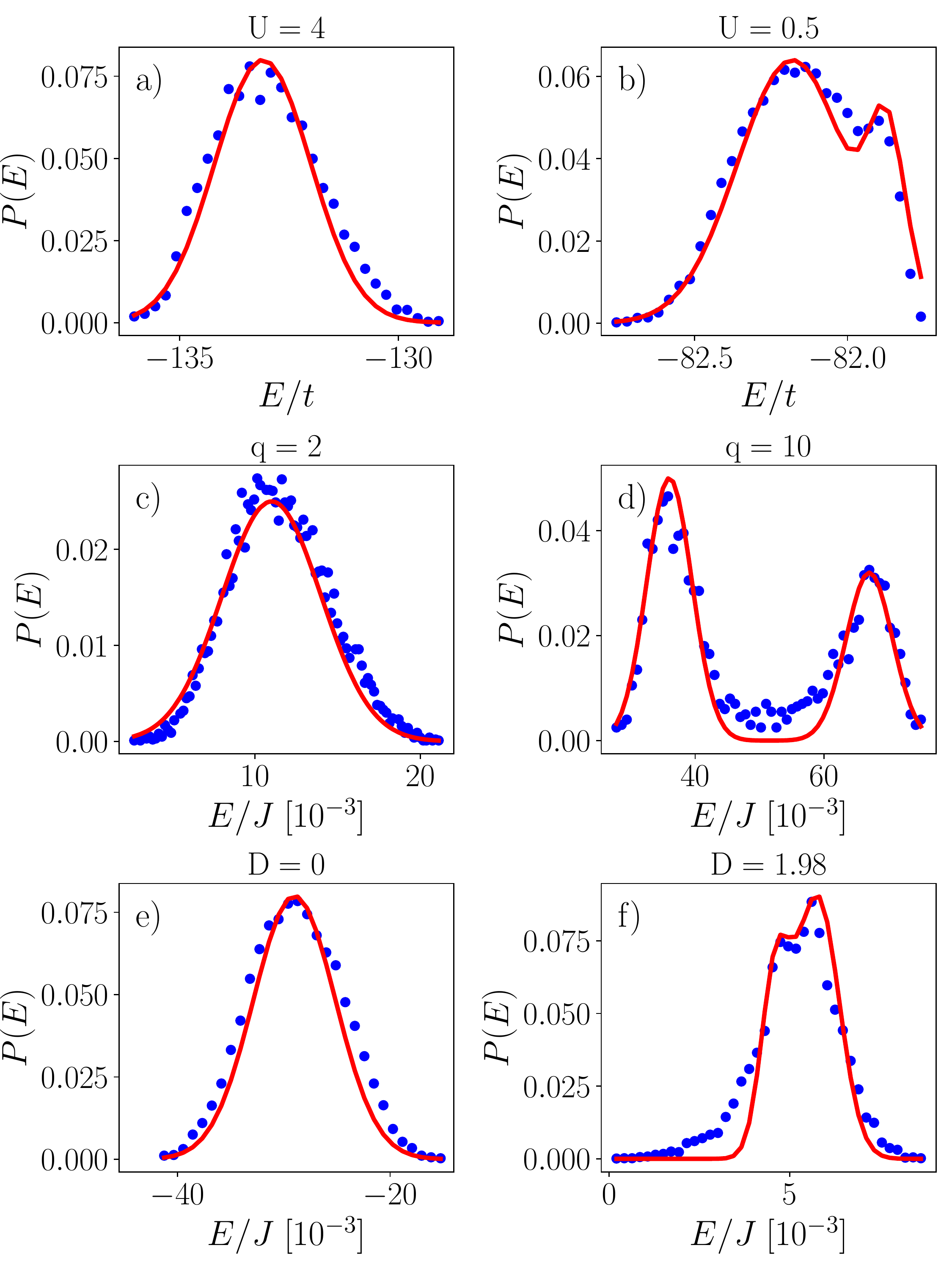}
    \caption{Probability  distribution  of  energy  at  temperatures close to the critical temperature for the  FK model (top row), $q$P model  (middle row) and the   BC model (bottom row). Plots in the left (right) column correspond to the continuous (discontinuous) phase transition. Points represent the MC data, lines are fits to the data.  }
    \label{fig:histograms}
\end{figure}

The ``standard'' method used to numerically determine the order of a phase transition is based on the finite--size effects that occur in discontinuous phase transitions \cite{challa, binder_challa, landau, FERNANDEZ1992485, shumpei}. It uses an {\it anomaly} behavior of the energy distribution $P(E)$ in the discontinuous phase transition at the critical point $T_{c}$. MC simulations usually generate the Gaussian distribution of energy $P(E)$ at a given temperature $T$, whose width indicates the value of specific heat $C_{v}(T)$. However, this is only entirely true for the continuous phase transition. When it comes to the discontinuous one, the situation is different. A single Gaussian distribution is observed only away from $T_{c}$. In the vicinity of a phase transition, $P(E)$ can be approximated by a superposition of two weighted Gaussians centered at energies $E_{-}$ and $E_{+}$, corresponding to two coexisting phases \cite{fk_maska}.   Therefore, an insightful study of the energy distribution can serve as a hint of whether one deals with continuous or discontinuous phase transitions. Figure~\ref{fig:histograms} shows the distributions obtained from our MC calculations for the three models studied in this work. In the left (right) column, energy distributions for a continuous (discontinuous)  phase transition are presented.  The existence of two Gaussians in Fig. \ref{fig:histograms} b), d), and f) corresponding to discontinuous phase transitions is evident.

To make our analysis more qualitative, we also estimate the order of the phase transition using the Binder--Challa--Landau cumulant \cite{challa, binder_challa} $V_{L}$ defined as
\begin{equation}\label{vl}
    V_{L} = 1 - \frac{\left < E^4 \right>_L}{3\left < E^2 \right>^2_L }.
\end{equation}
This quantity is known to have slightly different behavior for different types of phase transitions~\cite{challa, shumpei, Lima2018FourthOrderBC, Selke2006, Murtazaev2011}. When a transition is continuous, a small minimum appears in $V_{L}$ at the transition point, which is associated with the peak in the specific heat. The minimum scales with the size of the system $L$ and disappears when $L\to\infty$. On the contrary, when the transition is discontinuous, the minimum in $V_{L}$ survives in the thermodynamic limit.
\begin{figure}
    \centering
    \includegraphics[width=0.95\columnwidth]{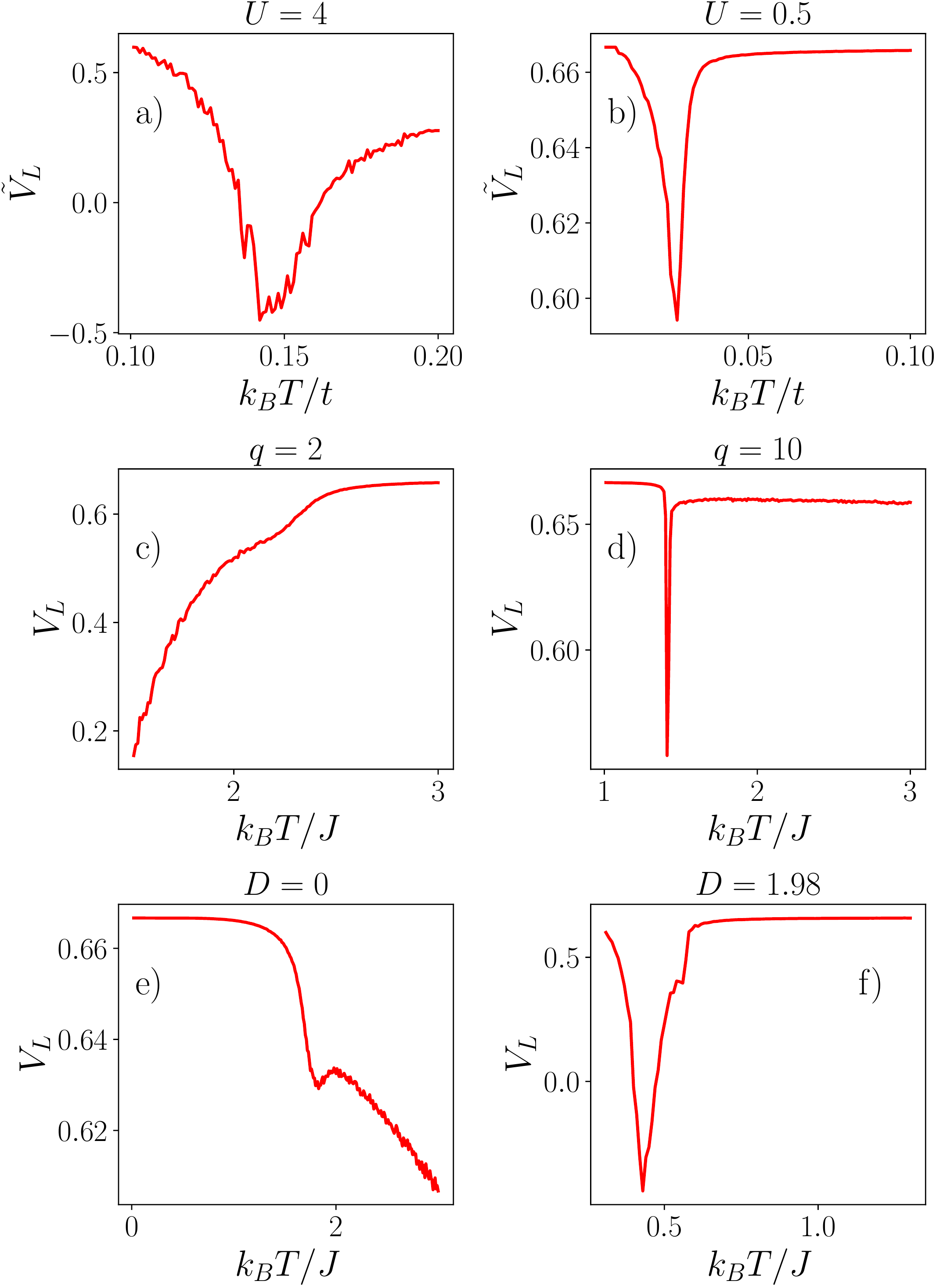}
    \caption{The Binder–Challa–Landau cumulant $V_L$ plotted as a function of temperature for the FK  model (top row), the \textit{q}P model (middle row), and the BC model (bottom row). The plots in the left (right) column correspond to the first--order (second-order) phase transition. In the case of the FK model, peaks in the cumulant are of order $10^{-6}$ to $10^{-4}$ and therefore we present $\tilde{V}_L=(V_L-0.6666)\times 10^{4}$ on the vertical axis in panels a) and b).
    } 
    \label{fig:vl_all}
\end{figure}
Fig. \ref{fig:vl_all} shows $V_{L}$ for the three models in the vicinity of the phase transition. One can see that for the BC and \textit{q}P models $V_{L}$ behaves as described above, while in the case of the FK model (Fig.~\ref{fig:vl_all}b), the effect is extremely weak. The relative change of the cumulant at the phase transition is of the order of $10^{-4}$ for the continuous phase transition and $10^{-6}$ for the discontinuous transition. Thus, in this case, the cumulant method cannot be used to determine the kind of phase transition. The difference between the FK and the two other models is that while the \textit{q}P and BC models are classical, in the FK model quantum degrees of freedom are coupled to classical degrees of freedom. The phase transition takes place in both subsystems at the same temperature (the charge susceptibility for the localized particles and the specific heat, that is calculated for the fermions, have peaks at the same temperature \cite{fk_maska}), but there are two different energy scales in this model. 
The first one is defined by an effective, fermion--mediated interaction between the localized particles driving a phase transition. The other, defined by the Fermi energy and the coupling to the classical particles, determines the behavior of the fermions and is significantly higher than $k_{\rm B}T_c$. The cumulant is calculated from fluctuations in the energy of the fermions. Because such fluctuations are weakly affected by the phase transition, its appearance is hardly visible in the temperature dependence of $V_L$.

The cumulant method also has some limitations for classical models.  The distinction between weak first--order transitions and second--order transitions, due to the large correlation length, requires long MC simulations on large lattice sizes \cite{FERNANDEZ1992485, shumpei, landau}. It is therefore desirable to find another method that would allow one to determine the nature of phase transition, even on relatively small lattices. This is particularly important for models such as the FK model, where this method does not work.

\section{Results}\label{sec:results}

\begin{figure}
    \centering
    \includegraphics[width=0.99\columnwidth]{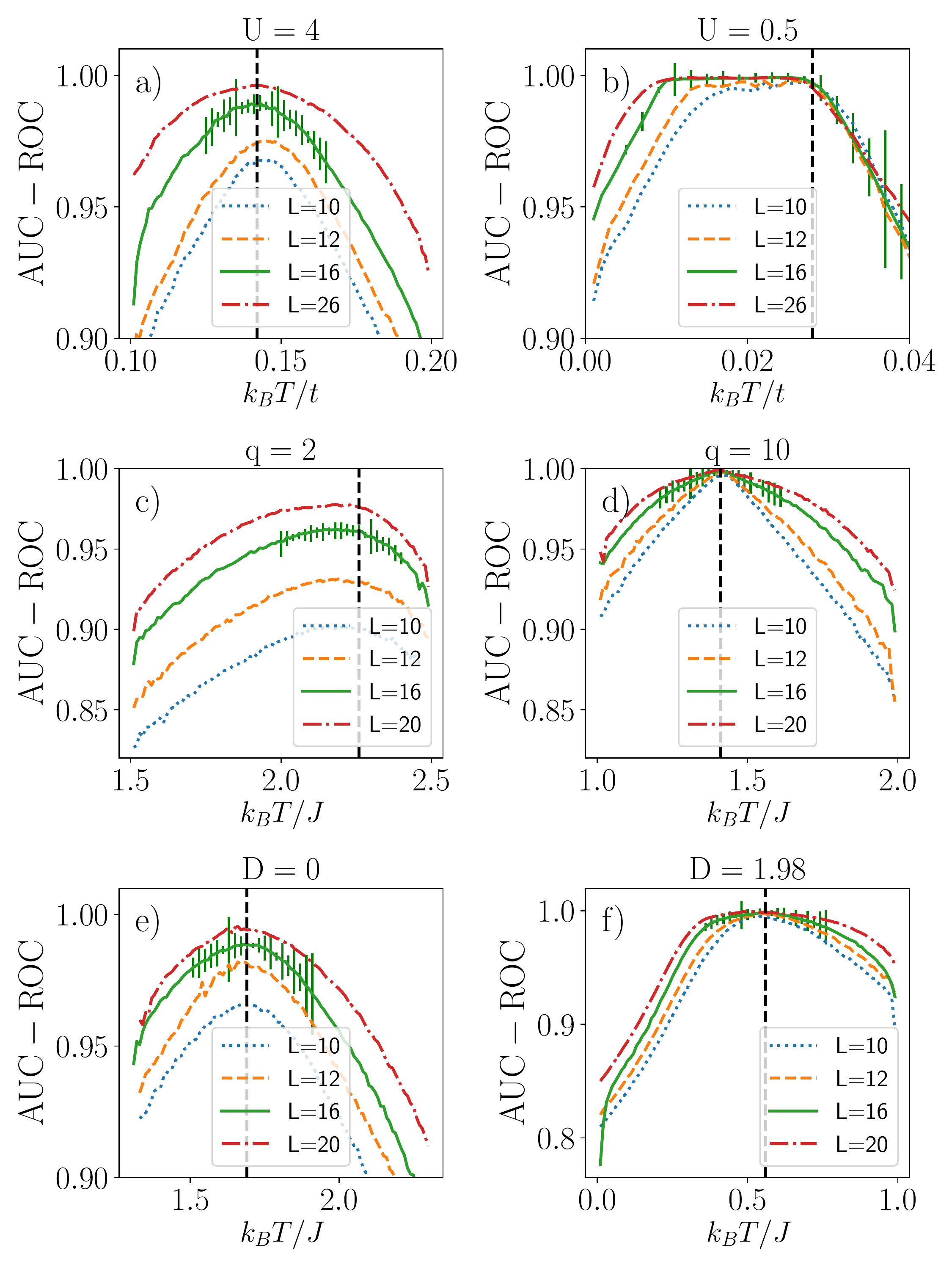}
    \caption{AUC--ROC versus temperature for continuous (left column) and discontinuous (right column) phase transitions: a)  the FK model, c) the \textit{q}P model, e) the BC model; b), d), f) - same as a), c), e), respectively, but for the discontinuous phase transition. Vertical black dashed lines depict the critical temperature extracted from MC simulations. The green bars represent statistical errors calculated for $16\times 16$ systems. For the sake of visibility their magnitude was multiplied by 10.}
    \label{fig:roc_curves}
\end{figure}

Figure~\ref{fig:roc_curves} summarizes the main result of this paper. The AUC-ROC curves are plotted versus temperature for model parameters corresponding to both continuous (left column) and discontinuous (right column) phase transitions. For all models, the accuracy has a W--like shape typical for the LbC method in the wide temperature range, but we show only the vicinity of the central maximum, which is essential for our discussion.
To verify how the results depend on the size of the system, we performed simulations for $L = 10, \ldots , 26$. Two main effects can be observed. For the continuous phase transition, the shape of the AUC-ROC curve depends only weakly on the size of the system, but the maximum value of the accuracy at $T_c$ converges to one as the size increases. On the contrary, for the discontinuous phase transition, the maximum value of the accuracy at $T_c$ is almost size independent, while the shape of the AUC-ROC curve changes significantly. Importantly, as the system increases, for all models studied, the maximum becomes less and less pronounced. In Fig.~\ref{fig:roc_curves} we also present error bars, which were calculated for a $16\times 16$ system. The magnitude of the uncertainty clearly indicates that for the first--order phase transitions the tendency to flatten the maximum is not due to statistical errors.
The finite size scaling and the method by which the uncertainty was determined are discussed in Appendix \ref{sec:fss}.

One can see in Fig.~\ref{fig:roc_curves} that, for continuous phase transitions, the accuracy always has the maximum exactly at the critical temperature, as predicted by the LbC method and our theoretical model. In the case of discontinuous phase transitions [Fig.~\ref{fig:roc_curves}b), d), f)]  the result depends on the system. For the BC and \textit{q}P models, the maxima of AUC-ROC align with the true $T_c$ extracted from MC simulations, but are softer than for continuous phase transitions, in agreement with the theoretical model. 
A strikingly different is the result for the FK Hamiltonian shown in Fig.~\ref{fig:roc_curves}b). Instead of a soft maximum, the AUC-ROC has a large plateau located at temperatures lower than and equal to $T_c$. The plateau spreads over a temperature range much wider than that of the region of phase coexistence. Additionally, this range increases as the size of the system increases. Therefore, its presence cannot be explained by the additional \textit{confusion} induced by the coexistence, and the toy model is not sufficient to explain its presence. 

To explain the origin of the plateau, we analyze the difference between the ion configurations generated by the MC simulations below and above $T_c$. 
\begin{figure}[ht]
    \centering
    \includegraphics[width=\columnwidth]{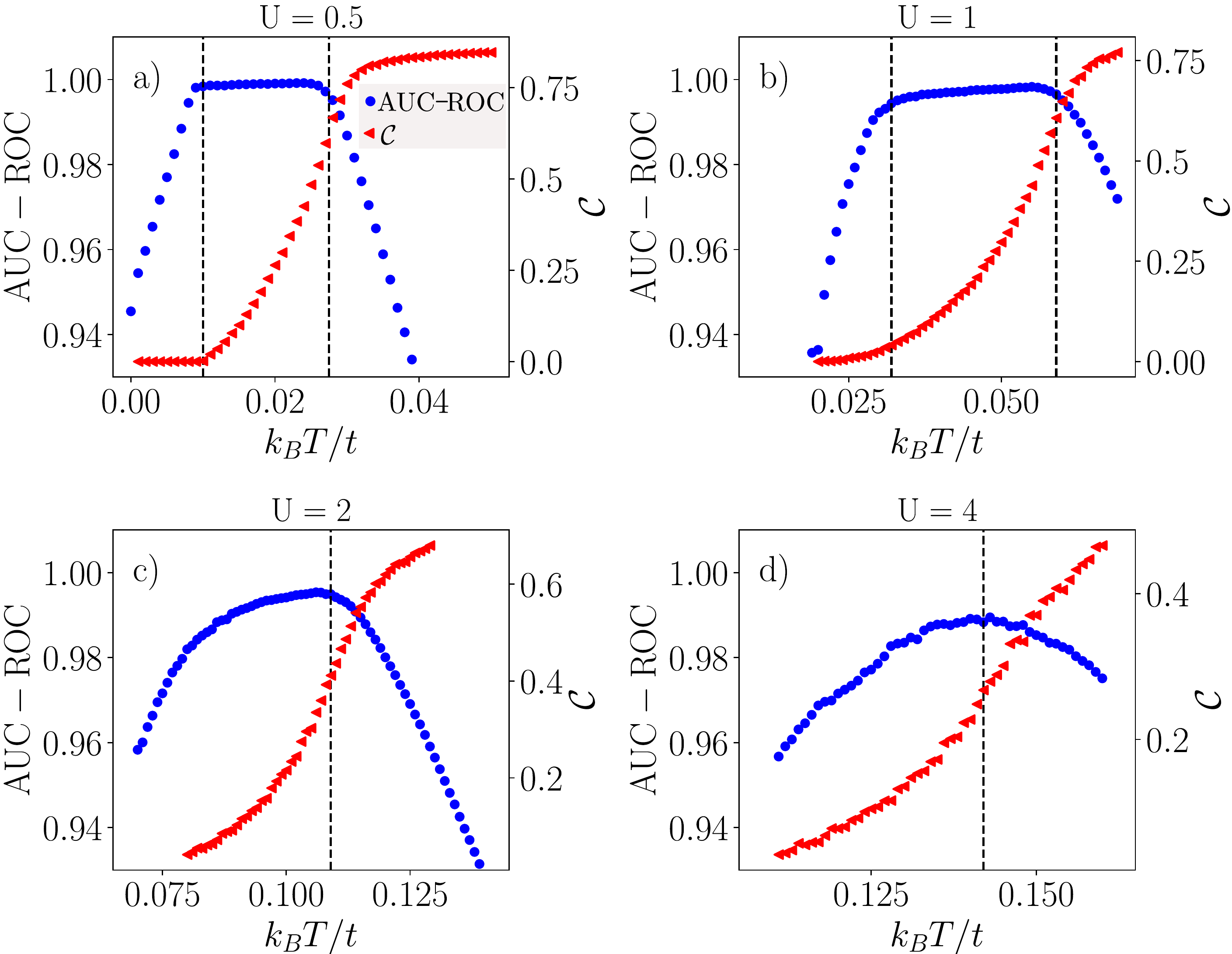}
    \caption{The AUC-ROC curve (red) and the concentration of defects $\mathcal{C}$ (blue dots) versus temperature for the FK model and for different values of $U$: a) $U=0.5$; b) $U=1$; c) $U=2$, d) $U=4$. A defect is defined as a deviation of the lattice site from the configuration corresponding to the checkerboard pattern of the entire lattice. For $U=0.5$  the phase transition is discontinuous and changes to continuous for $U\approx 1$.
    \label{fig:defects}}
\end{figure}
Figure \ref{fig:defects} shows the temperature dependence of the concentration of defects ${\cal C}$ in a perfect checkerboard order of the ions (ground state configuration of the FK model) and the corresponding AUC-ROC. Here, a single defect is defined as the deviation of a lattice site from the configuration corresponding to the checkerboard pattern, as illustrated in Appendix \ref{defect_defs}. Different panels show results for different values of the interaction $U$. For interaction $U=0.5$, which leads to a discontinuous phase transition, ${\cal C}(T)$ for $T\lesssim T_c$ is much steeper than for $U=4$, for which the system undergoes a continuous phase transition.
By fitting a linear function to regions of growth of ${\cal C}(T)$ in Figs. \ref{fig:defects} a) and d), we found that the slope of ${\cal C}(T)$ for $U=0.5$ is approximately five times greater than for $U=4$ \footnote{The differences in temperature dependence of configurations in the FK model in the weak and strong interaction regime has been discussed in Ref.~\cite{MMKC2005}.}.
This means that differences between configurations generated at two close temperatures are much more pronounced for a discontinuous phase transition than for a continuous one.

\begin{figure}[ht]
    \centering
    \includegraphics[width=0.99\columnwidth]{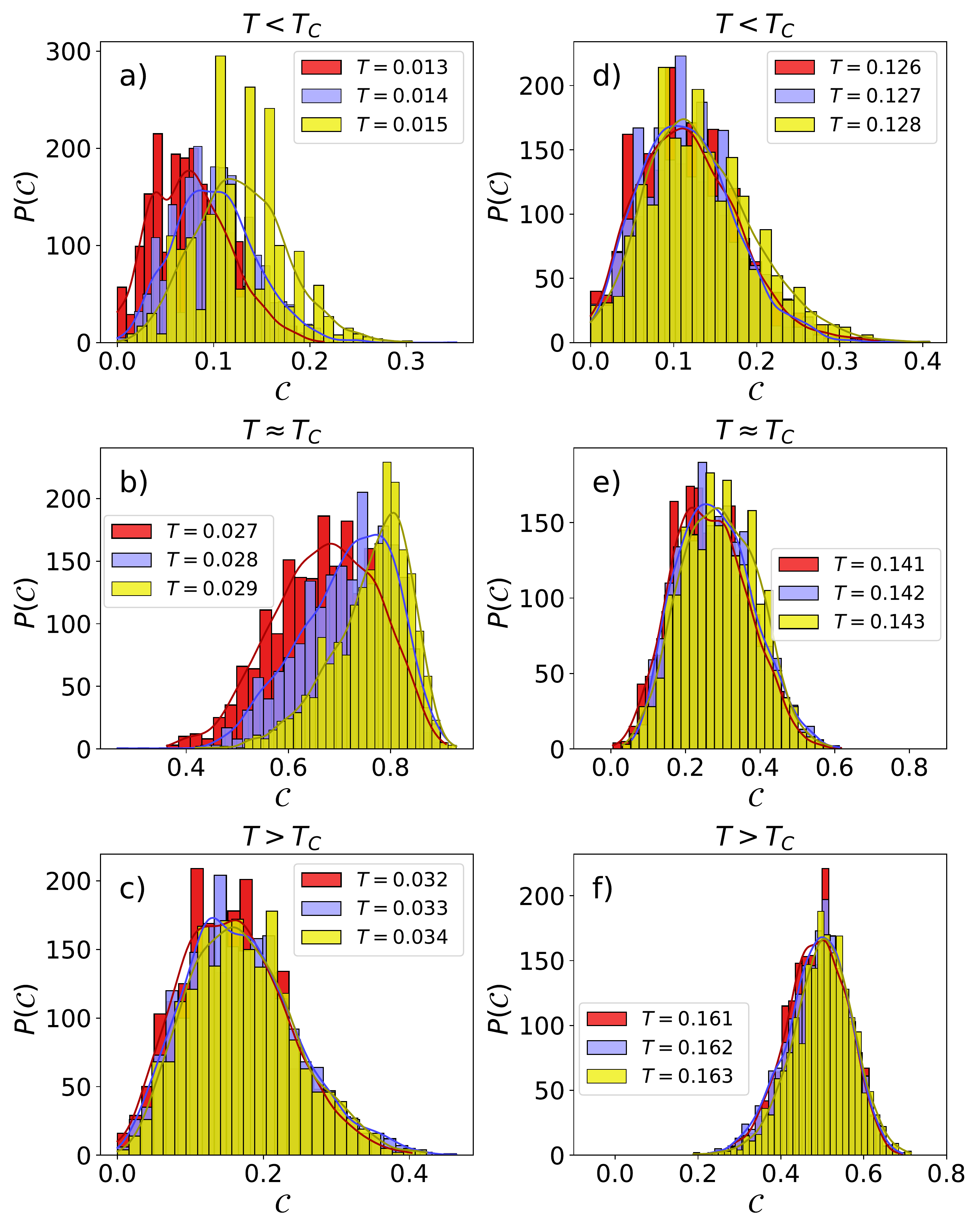}
    \caption{Probability distribution of defect concentration $\mathcal{C}$ for the FK model at different temperatures. Left column: for a continuous phase transition, $U=4$; right column: for a discontinuous phase transition, $U=0.5$.}
    \label{fig:probablility}
\end{figure}

 This explanation is supported by the results presented in Fig.~\ref{fig:probablility}, where we show the distributions of the defect concentration $P(\mathcal{C})$  at three close temperatures. When the system undergoes a continuous phase transition (panels a, b, and c) and when $T>T_c$ for a discontinuous phase transition (panel f), $P(\mathcal{C})$ are almost identical for such small temperature increments. On the contrary, for $U=0.5$ and $T \lesssim T_c$ (panels d and e), the distributions for temperatures that differ by $\delta T=0.001$ are significantly different. We argue that this difference along with the specific temperature dependence of $P(\mathcal{C})$ creates sufficient conditions for the neural network to recognize configurations from the plateau region as qualitatively different and thus ``learn'' the concentration of defects.

\begin{figure}
    \centering
    \includegraphics[width=0.99\columnwidth]{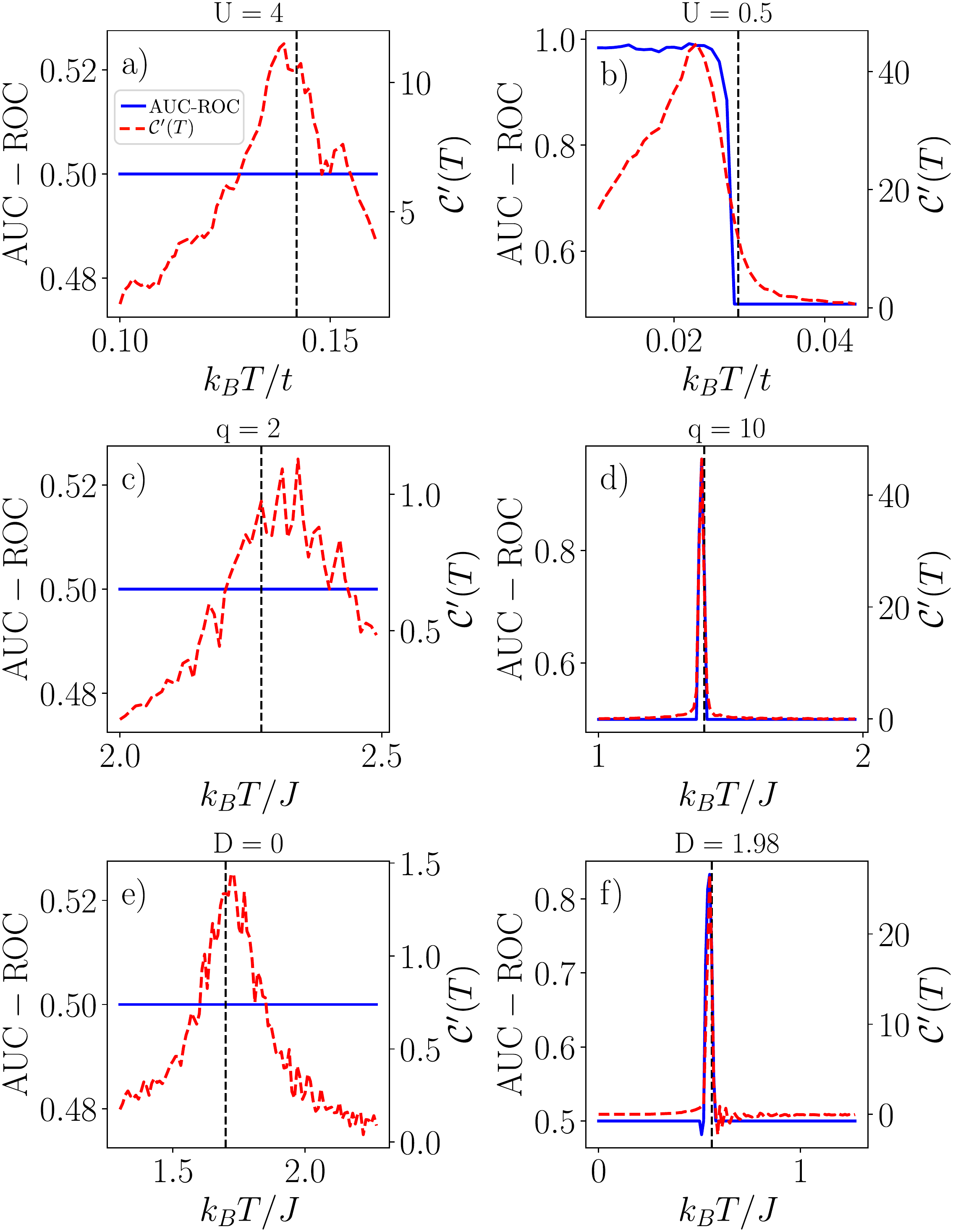}
    \caption{Qualitative differences between configurations at two close temperatures plotted as a function of temperature $T$: derivative of defect concentration  $\Delta {\cal C'}(T)$ (red dashed line) and the corresponding neural network accuracy AUC-ROC (solid blue line) of classifying the configurations from the temperatures $T$ and $T+\delta T$ as belonging to two different phases. Panels a), c), and e) present data for discontinuous phase transition, while b), d), and f) for continuous phase transitions, for the F-K, Potts, BC models respectively. The vertical dashed line depicts $T_c$ extracted from MC simulations. }
    \label{fig:roc_def_dT}
\end{figure}
\begin{figure}
    \centering
    \includegraphics[width=0.98\columnwidth]{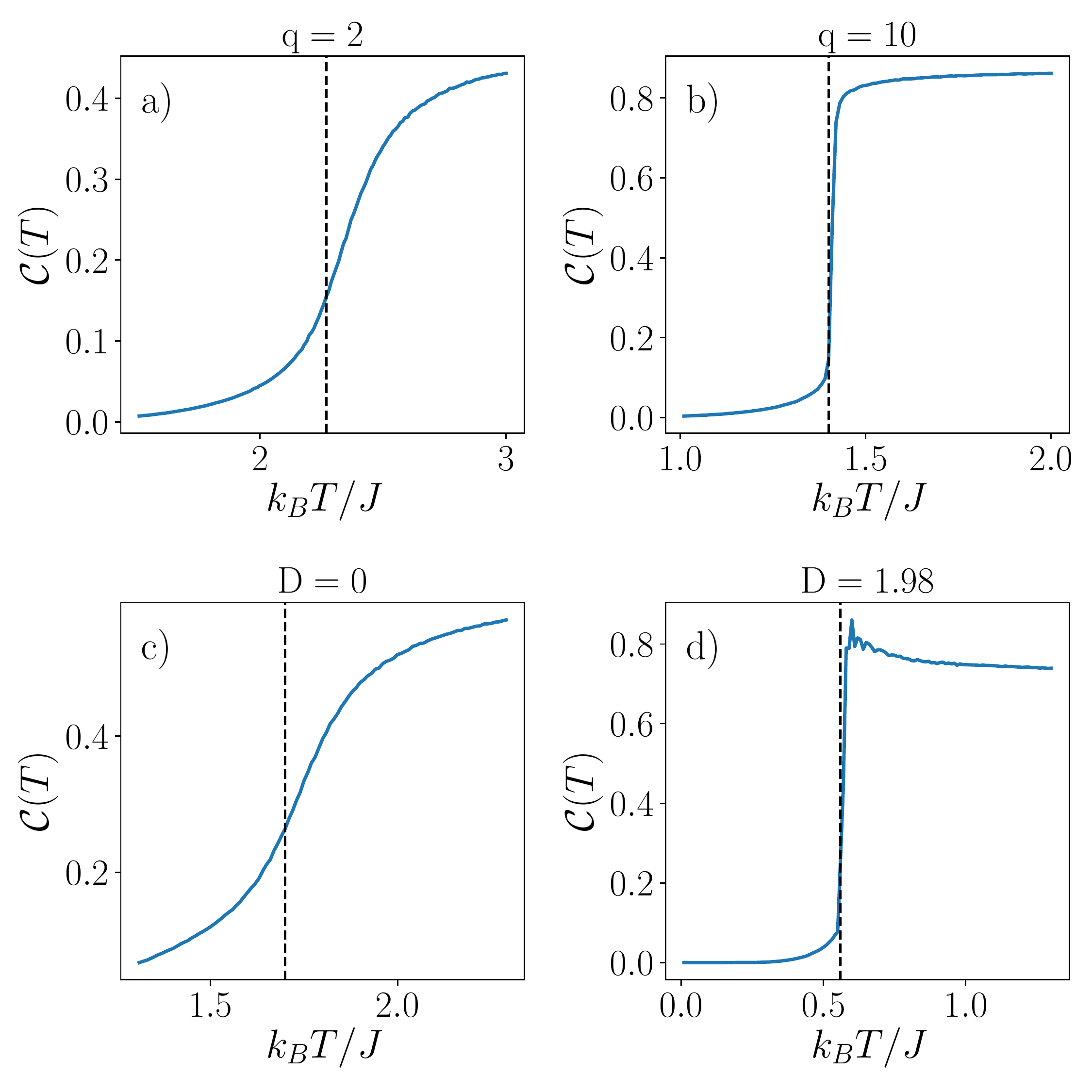}
    \caption{Concentration of defects $\mathcal{C}(T)$ as a function of temperature for continuous phase transitions for the \textit{q}P model (a) and the BC model (c). Panels b) and d) show the same as panels a) and c), respectively, but for the discontinuous phase transition. The vertical lines depict the critical temperature determined in MC simulations.}
    \label{fig:defects_all}
\end{figure}

To confirm or refute this conjecture, we performed explicit tests of the ability of a neural network to identify different concentrations of defects. To do that, we trained the neural network with configurations generated at only two very close temperatures $T$ and $T+\delta T$, labeling them as low- and high-temperature phases, respectively. We used $\delta T=0.005\, k_{\rm B}T/t$ for the FK model and $\delta T=0.02\, k_{\rm B}T/J$ for the BC and Potts models. Then, sweeping $T$ in a broad range, we calculated the effectiveness of the network in distinguishing between the configurations generated at $T$ and $T+\delta T$, and checked how this effectiveness depends on the slope of ${\cal C}(T)$. The results are presented in Fig. \ref{fig:roc_def_dT}, where AUC-ROC is compared to the derivative of ${\cal C}(T)$ [approximated as ${\cal C}'(T)\approx \Delta{\cal C}(T)/\delta T$, where $\Delta {\cal C}(T)={\cal C}(T+\delta T)-{\cal C}(T)$]. It can be seen that the AUC-ROC is exactly 0.5, which is the accuracy of a random classifier, unless ${\cal C}'(T)$ is really large.
For instance, for second order phase transitions the AUC-ROC is equal to 0.5 in the whole temperature range, because the concentration of defects grows relatively slowly with temperature (see also Figs. \ref{fig:defects_all} a, c) and, as discussed above for the FK model, their distributions are indistinguishable. 
For discontinuous phase transitions, the AUC-ROC peaks around $T_c$, when ${\cal C}'(T)$ is above a certain threshold value (Fig. \ref{fig:defects_all} b, d, f). For the \textit{q}P and BC models, it occurs only in an extremely narrow temperature range around $T_c$ and is caused by an almost step-like increase in the defect concentration at $T_c$, shown in Figs. \ref{fig:roc_def_dT}d) and e). Thus, according to our argumentation, no plateau can be expected there. 
A plateau is observed for the FK model with $U=0.5$ (Fig. \ref{fig:roc_def_dT}b), within the same temperature range and with almost the same value of AUC-ROC $\approx 1$ as in Fig. \ref{fig:defects}, when the network was trained using the entire dataset. The plateau is visible when ${\cal C}'(T)\gtrsim 17$ and disappears when ${\cal C}'(T)$ falls below this value. This proves that the neural network classifies the configurations from the temperature range in which the plateau exists, as qualitatively different (low- and high-temperature ones, for example), even if the belong to the same low temperature phase.

In Appendix \ref{sec:fss} we show that this behavior becomes even more visible as the system size increases. In particular, the temperature range where the plateau occurs for the FK model becomes wider, and the value of AUC-ROC increases. This is related to the change in the shape of ${\cal C}(T)$ with the size shown in Fig.~\ref{fig:C_N}a). While the slope increases only slightly, the region of fast increase of the defect concentration moves towards lower temperatures. 
As can be seen in the left column in Fig. \ref{fig:roc_curves}, the AUC-ROC also increases for the continuous phase transition, which can be attributed to the decrease of the width of the phase transition and steeper ${\cal C}(T)$, as for the FK model can be seen in Fig.~\ref{fig:C_N}b). This relation between network performance and ${\cal C}(T)$ for different system sizes is an additional argument that supports the idea of the network's ability to learn the defect concentration.


\section{Summary}\label{sec:summary}
In this paper, we have applied Monte Carlo calculations and the Learning by Confusion machine learning method to study phase transitions in three microscopic models. 
We demonstrated that the efficiency of the LbC method in recognizing the kind of phase transition strongly depends on the character of the transition. In particular, we have shown that ({\it i}) in the case of a discontinuous phase transition, this method cannot always be used to determine the critical temperature, and ({\it ii}) the shape of the LbC accuracy potentially can be used to identify a discontinuous phase transition.
This behavior was predicted with the help of a simple model that describes spin configurations close to continuous and discontinuous phase transitions.


Our results show that for the BC and \textit{q}P models, the performance of LbC in recognizing the kind of phase transitions is weak, in some cases insufficient for a direct practical application. Although the distinction between first-- and second--order transitions increases as the system size increases, the required size can make calculations unfeasible. However, for these models, the standard cumulant method should be applied.
For the FK model, the neural network is able to capture local differences between configurations in the low-temperature phase with almost perfect accuracy. This ability leads to a qualitatively different shape of overall classification accuracy, making the LbC method an efficient tool for discriminating between continuous and discontinuous phase transitions in this model. It is an important result because the Binder–Challa–Landau cumulant, being a standard method used in this context, performs very badly for the FK model. 

While training the neural network requires additional computational time and memory, there are advantages to using the proposed approach, especially for the FK model. The most obvious one is that it is sufficient to train the network only once for a given model. The trained network can then be used for other model parameters.
In addition, for a convolutional neural network, the number of operations increases almost linearly with the size of the system, which makes the finite-size scaling relatively easy, provided the configurations are already generated. 
\section*{Acknowledgments}
 This work was supported by the National Science Centre (Poland) under grant DEC-2018/29/B/ST3/01892. Numerical calculations have been carried out using High Performance Computing resources provided by the Wrocław Centre for Networking and Supercomputing. We would like to thank Jaros{\l}aw Paw{\l}owski for his help with the high-performance calculations.
\appendix

\section{Finite size scaling and statistical errors}\label{sec:fss}



Figure \ref{fig:roc_curves} shows the temperature dependence of AUC-ROC for different sizes of the systems. When approaching the thermodynamic limit, the behavior of the system close to $T_c$ changes from a crossover to a true phase transition. Peaks in specific heat and magnetic susceptibility become narrower, and so does the temperature region where the critical fluctuations occur.  

\begin{figure}[h]
    \centering
    \includegraphics[scale=0.4]{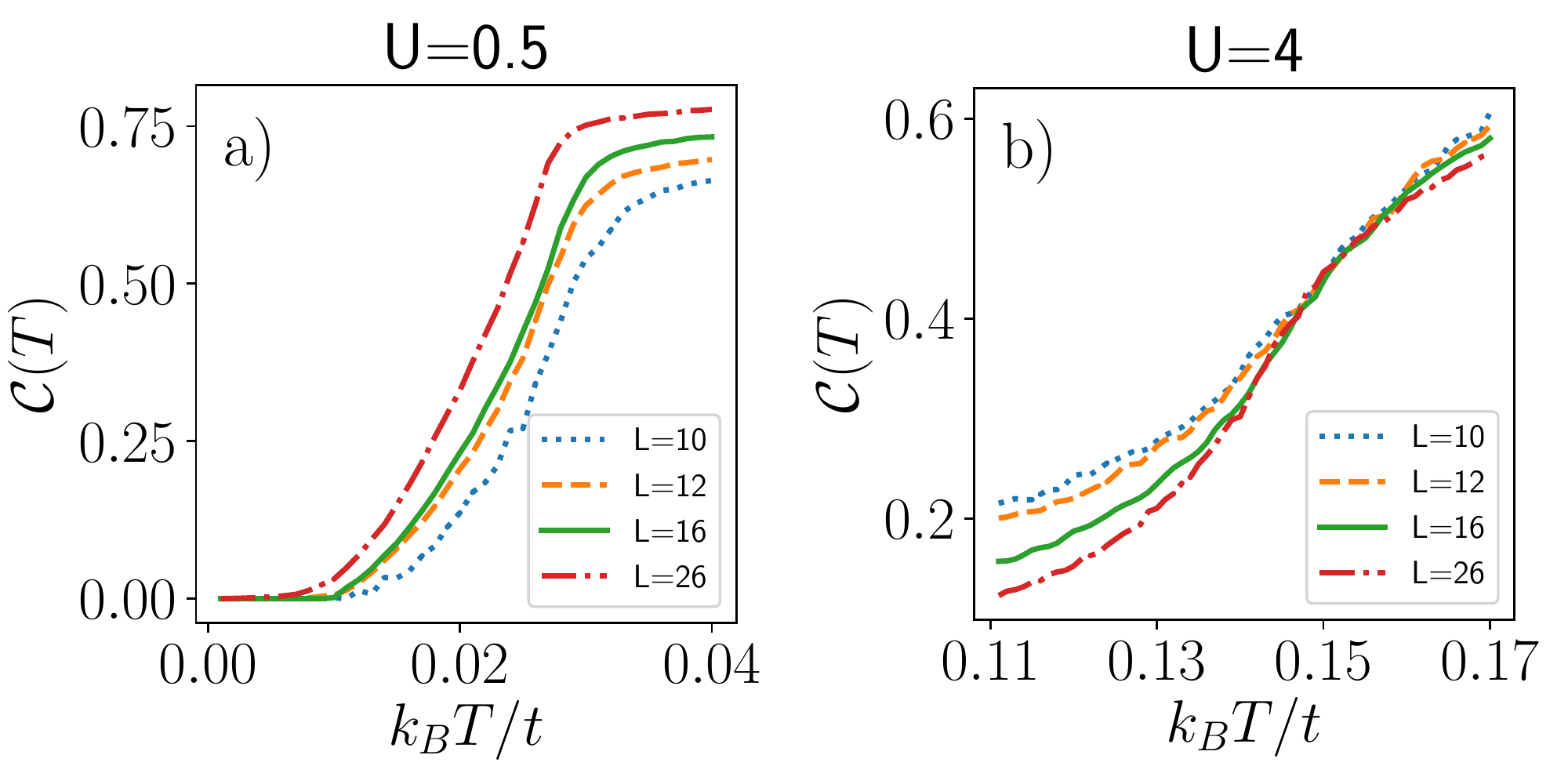}
    \caption{${\cal C}(T)$ for the FK on lattices of different sizes for $U=0.5t$ (a) and $U=4t$ (b).}
    \label{fig:C_N}
\end{figure}

To check whether these changes in network performance are consistent with our conjecture of the network's ability to identify defect concentrations, we calculated how ${\cal C}(T)$ for the FK model is affected by the size of the system. The results are presented in Fig.~\ref{fig:C_N}. One can see there that these results indeed correspond to those presented in Fig.~\ref{fig:roc_curves}. For $U=0.5t$ the region of the fastest increase in ${\cal C}(T)$ shifts towards lower temperatures, leading to a widening of the plateau. 
For $U=4t$ ${\cal C}(T)$ becomes steeper around $T_c$ and thus the network performance increases. 

Finite size effects are one potential source of unreliability of the results obtained. Another is the uncertainty of the results predicted by the neural network. To estimate its magnitude, we used a method proposed in Ref.~\cite{pmlr-v48-gal16}. It has been shown there that a neural network with arbitrary depth and non-linearities, with dropout applied before every weight layer, is mathematically equivalent to an approximation
to the probabilistic deep Gaussian process. Then, the classification uncertainty can be obtained by collecting the results of stochastic forward passes through the network with an additional dropout layer. The error bars presented in Fig.~\ref{fig:roc_curves} were obtained for dropout with probability $p=0.5$.

Apart from calculating the errorbars, to make our predictions more confident, for each presumable critical temperature $T'_{C}$, the accuracy of the neural network prediction was validated on a randomly chosen testing dataset (random split in training and testing datasets in proportion 80\% to 20\%).  However, in any case, we did not observe that our model overfits - the difference between AUC-ROC received for the training and testing datasets were of the order of 0.01-0.05\%.  It further confirms the reliability of the proposed model.

\section{Computational details\label{app:details}} 
The input datasets are generated using Metropolis Monte Carlo simulations. The BC model is simulated with the standard MC with the local update method. For the \textit{q}P model, in order to accelerate the calculations, we used the Wolff cluster algorithm~\cite{wolff}. In the case of the FK model, the configurations are computed by applying a modified version of the classical Metropolis algorithm, where the interaction energy is replaced by the free energy obtained as a result of numerical diagonalization of the Hamiltonian for a given ionic configuration. More details about this method can be found in ~\cite{fk_maska}.

Each of the models mentioned above is implemented on a $16\times 16$ square lattice with periodic boundary conditions. At a given temperature, the system is first relaxed to reach thermal equilibrium, and then 10000 snapshots of the system states are taken in step intervals adjusted to the autocorrelation length.

\begin{figure}
    \centering
    \includegraphics[width=0.43\textwidth]{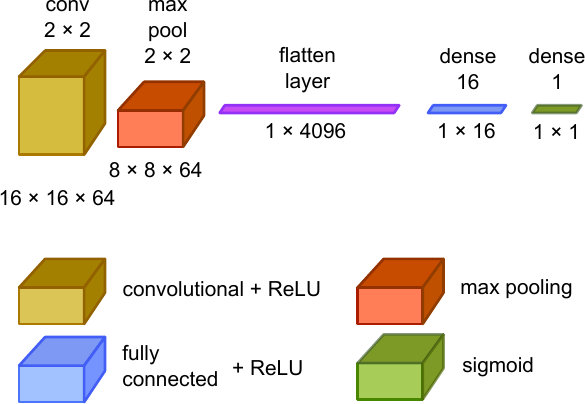}
    \caption{Architecture of the neural network used during the LbC scheme analysis. The net consists of one convolutional layer with 256 $2\times 2$ filters, one subsampling (\texttt{max pool 2x2}) layer and two fully connected layers with ReLU activation function. The output layer is activated with the sigmoid function given by Eq. \ref{eq:sigmoid}.}
    \label{fig:architecture}
\end{figure}

For all models studied, we use a neural network consisting of one convolution layer (CNN) with 256 filters of shape $2\times 2$, one max-pooling layer reducing twice the dimensionality of the MC configurations, and two fully connected layers at the end (Fig. \ref{fig:architecture}). 
The convolutional and first fully connected layers are activated with the ReLU function defined as
\begin{equation}
    \mathrm{ReLU} = \max(0, x).
\end{equation}
The output layer is composed of one neuron that is activated with the sigmoid function $\sigma(x)$,
\begin{equation}
    \sigma(x) = \frac{1}{1+e^{-x}},
    \label{eq:sigmoid}
\end{equation}
returning the probability of a configuration belonging to the high-temperature phase.

The network is trained with the Adam optimization algorithm~\cite{adam} with the optimal choice for the learning rate $\eta=0.001$. To reduce overfitting, we also apply, during the training process, the \texttt{L2} regularization penalty technique~\cite{l1l2} together with the \texttt{early-stopping} method~\cite{regularization}.

For a given temperature, $8000$ out of $10000$ statistically independent Monte Carlo configurations are used in the training, while the remaining $2000$ configurations are added to the testing set. Taking into account that each LbC computation requires configurations corresponding to $100$ different temperatures, the total number of configurations used within the training process is $800000$. Such a set is generated for each microscopic model under study. 
The batch size is set at $256$, which, according to empirical observations, gives us a good balance between the speed of the calculations and the stability of the results obtained.  

\begin{figure}
    \centering
    \includegraphics[width=0.5\columnwidth]{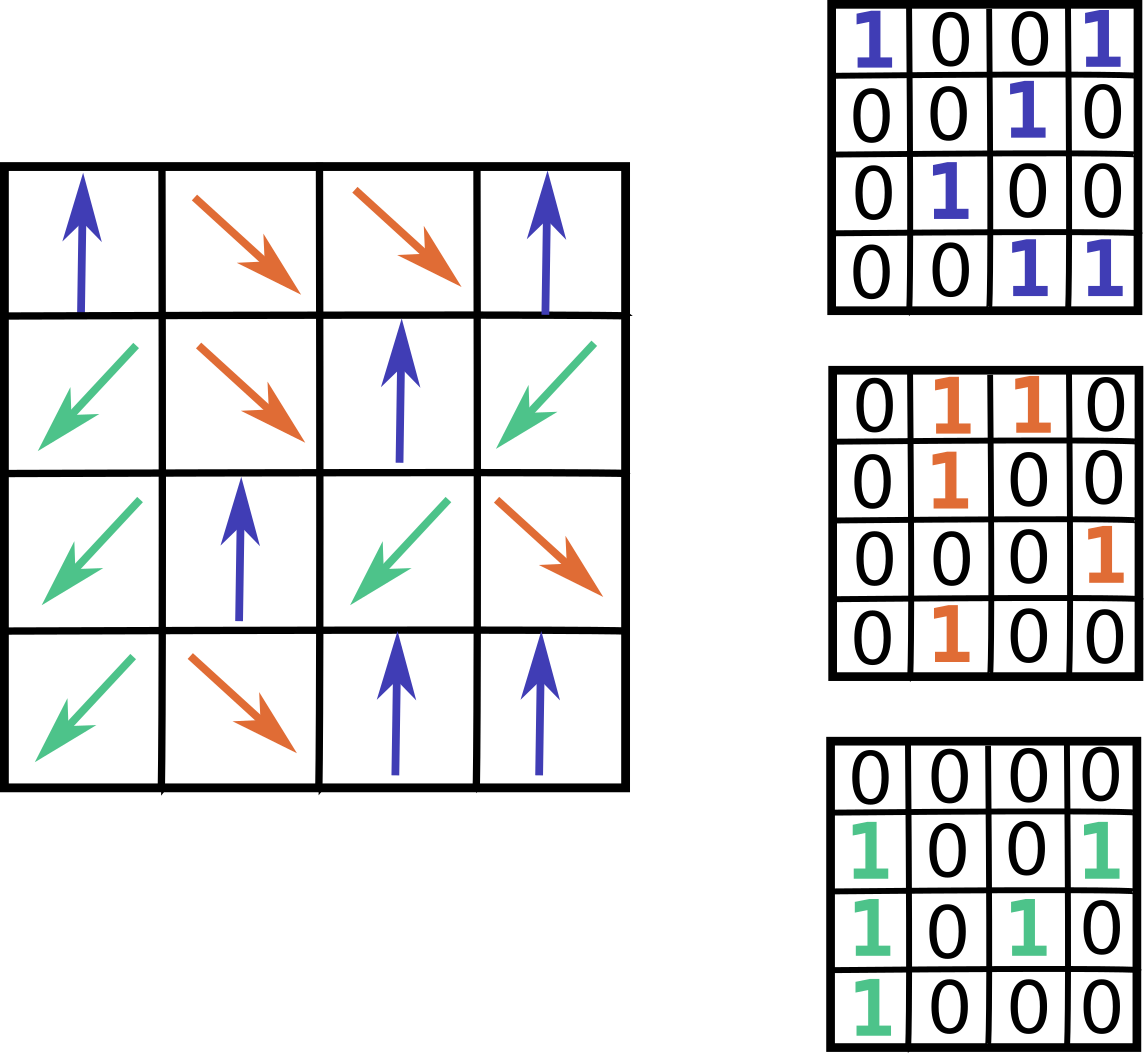}
    \caption{The transformation of the raw MC configurations before feeding them into the neural network. From each configuration of the size $N\times N$ $q$ new 'configurations' of the same size are created in a way analogous to the hot--encoding scheme, which is usually used for the labels' transformation. }
    \label{fig:input_neural}
\end{figure}

Although the aforementioned architecture of the {\color{blue}{Neural Network}} (NN), together with the preset hyperparameters, is used in all of our three models, there are some modifications to the input data for the neural network. In the case of the FK model, the configurations of shape $16\times 16$ are fed directly into the CNN layer. The situation is slightly more complicated when it comes to the \textit{q}P model. In this case, the number of input channels is equal to the value of parameter \textit{q} (see Sect.~\ref{subsec:Potts_model}), which corresponds to the number of ground states that occur in this model. Then, each channel is created on the basis positions of spins pointing in the same direction using a one-hot encoding scheme. Similar pre-processing of MC configurations has already been applied in Ref. \onlinecite{potts-neural}. An example illustrating this method on the $4\times 4$ lattice size is presented in Fig.~\ref{fig:input_neural} and can be trivially extended to the larger lattice sizes.

The transformation of the Monte Carlo configurations, obtained from the simulations of the BC model, is analogous to that performed in the case of the \textit{q}P model. The only difference is that, in the case of the BC model, independently of the model parameters, the spin at a given lattice site can always take three values: -1, 0, 1, so that each configuration always generates three channels.
The reason behind such a choice of data preparation is the stability of the results obtained from the neural network.

Neural network computations along with data preprocessing steps are performed using the \texttt{Python} programming language supported by packages \texttt{Keras}~\cite{chollet2015keras} and \texttt{Scikit-learn}~\cite{pedregosa2011scikit} packages.

\section{The AUC-ROC method\label{auc-roc}}

\begin{figure}[h]
    \centering
     \includegraphics[width=0.7\columnwidth]{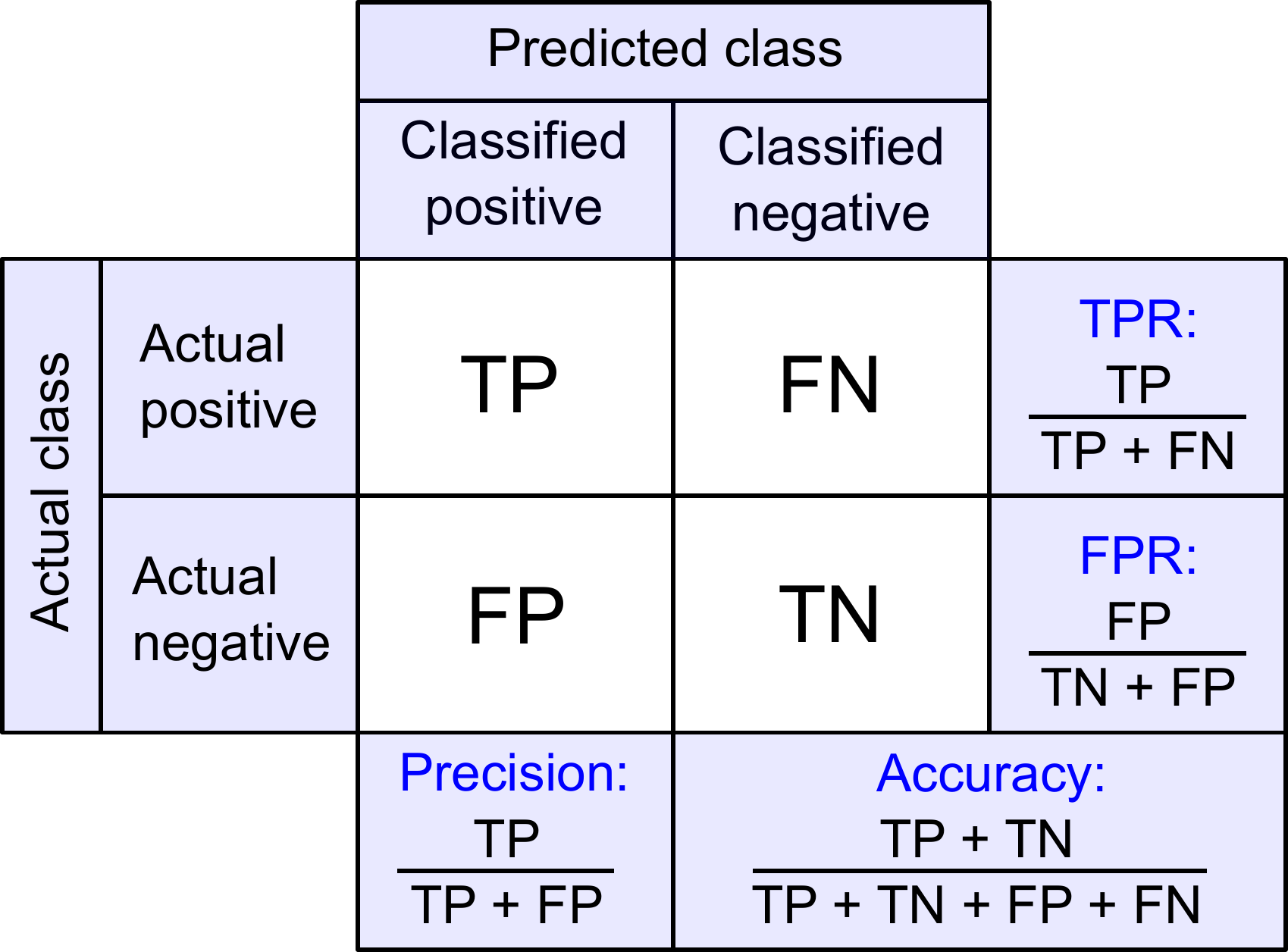}
    \caption{Schematics of the confusion matrix for the binary classification problem including definitions of basic terms used in the assessment of model's performance, i.e. precision, FPR (\emph{false positive rate}), accuracy, specificity and TPR (\emph{true positive rate}). Each cell of the matrix (rectangle) represents a number of  true and false classification predictions. }
    \label{fig:confusion_matrix}
\end{figure}

The AUC-ROC is an evaluation metric that represents the probability that the random pair of positive and negative samples is correctly classified ~\cite{ROC-AUC}. It requires the definition of a confusion matrix, a table with four cells, to which we assign a number of positively or negatively classified samples on the basis of their actual labels. Since the classification result can be false or true and positive or negative, there are four different possibilities (Fig. \ref{fig:confusion_matrix}).

From the confusion matrix, several additional metrics emerge. Two of them, that is, \emph{True Positive Rate} (TPR) and \emph{False Positive Rate} (FPR) are used in a definition of the ROC curve. TPR notifies about a fraction of a positive class that is correctly classified,

\begin{equation}
    \mathrm{TPR} = \frac{\mathrm{TP}}{\mathrm{TP} + \mathrm{FN}},
\end{equation}

while FPR,
\begin{equation}
    \mathrm{FPR} = \frac{\mathrm{FP}}{\mathrm{FP} + \mathrm{TN}},
\end{equation}
measures the fraction of the negative class incorrectly classified by a machine learning model.

\begin{figure}
    \centering
    \includegraphics[width=0.6\columnwidth]{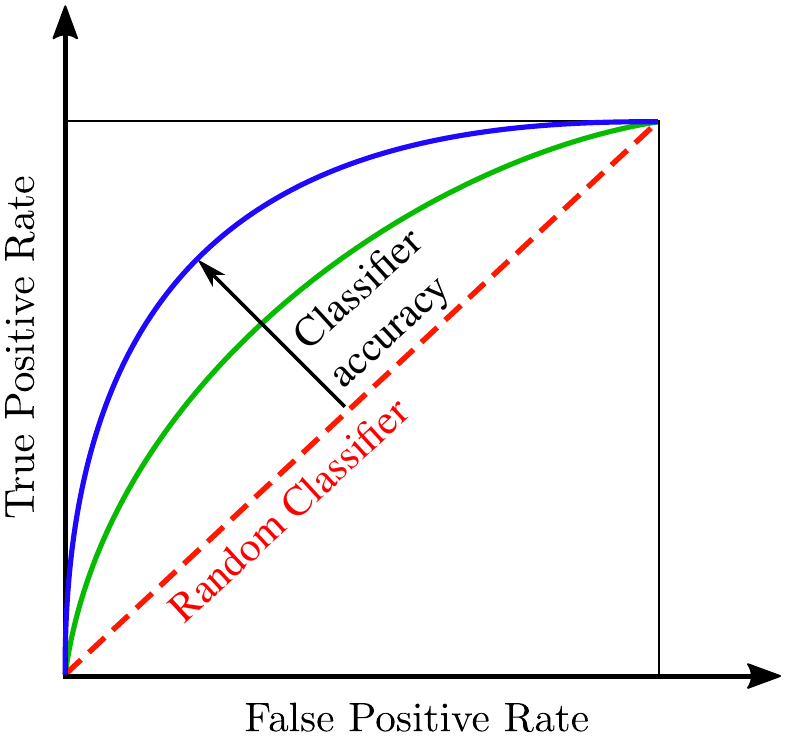}
    \caption{The ROC curve representing the function of TPR (\textit{True Positive Rate}) on FPR (\textit{False Positive Rate}) of a model. The diagonal line corresponds to  the case where a model classifies samples in a random manner. In general, the bigger is the function 'convexity', the larger is the model performance.}
    \label{fig:roc_curve}
\end{figure}

The ROC curve is created by plotting TPR against FPR 
for different discrimination thresholds.
The typical shapes of the curve are visualized in Fig. \ref{fig:roc_curve}. The diagonal line represents the case where a model classifies samples in a random manner; therefore, it does not have any discrimination capacity.

In general, the capability of a model to assign samples into right categories improves further when the curve is near the upper left corner of the plot. To quantitatively describe this fact, we use the \emph{Area Under the Curve} (AUC) evaluation metrics, which measures the integral under the ROC curve. The bigger this integral is, the better the model is at making predictions.

\begin{figure}
    \centering
    \includegraphics[scale=0.55]{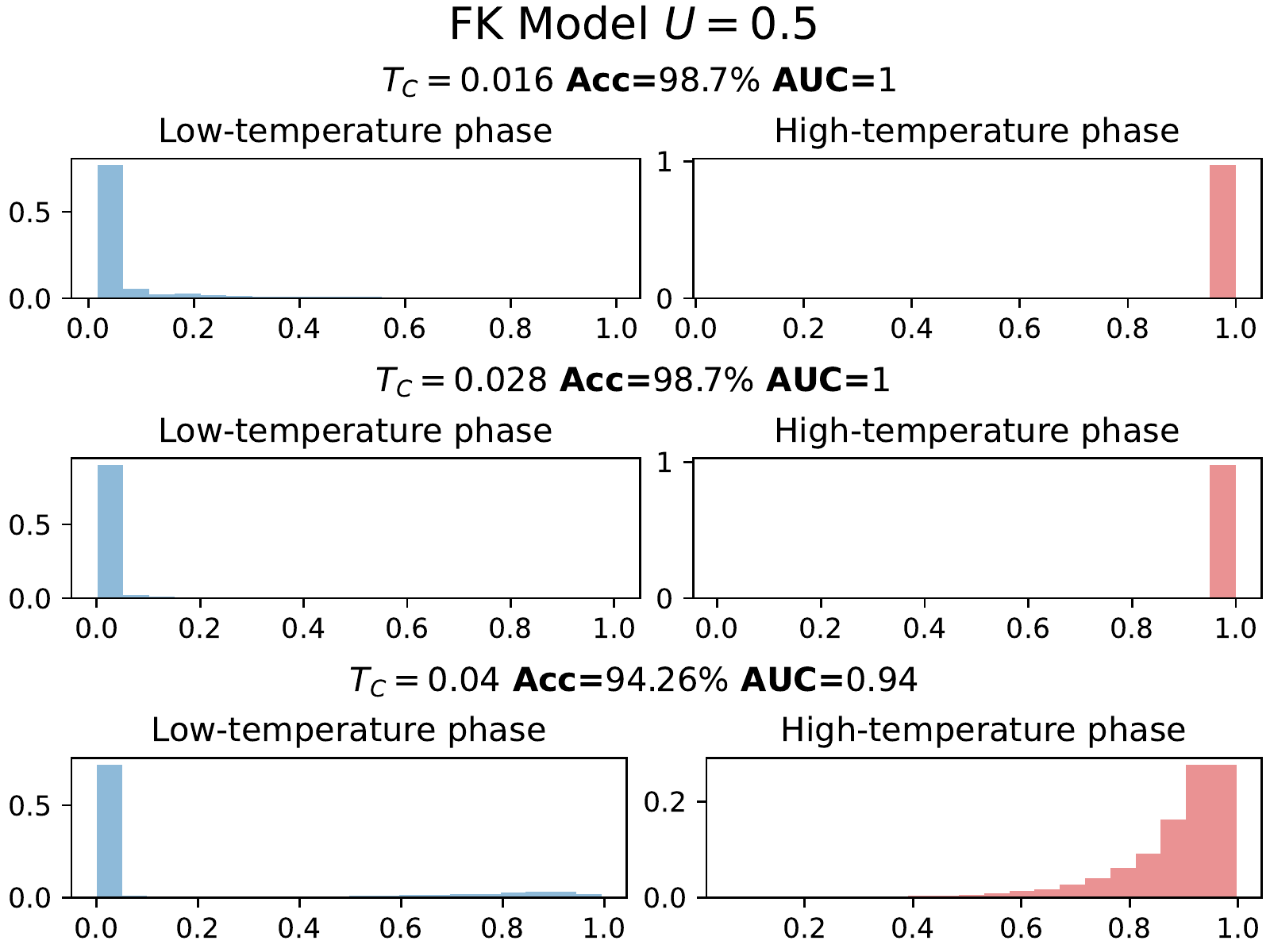}
    \caption{Distributions of outputs of the sigmoid activation function $\sigma(x)$ from the neural network fed with the spin configurations generated for FK model $U=0.5$ (discontinuous phase transition) in the case of different critical temperature $T'_{c}$.}
    \label{fig:U=0.5_auc}
\end{figure}

\begin{figure}
    \centering
    \includegraphics[scale=0.55]{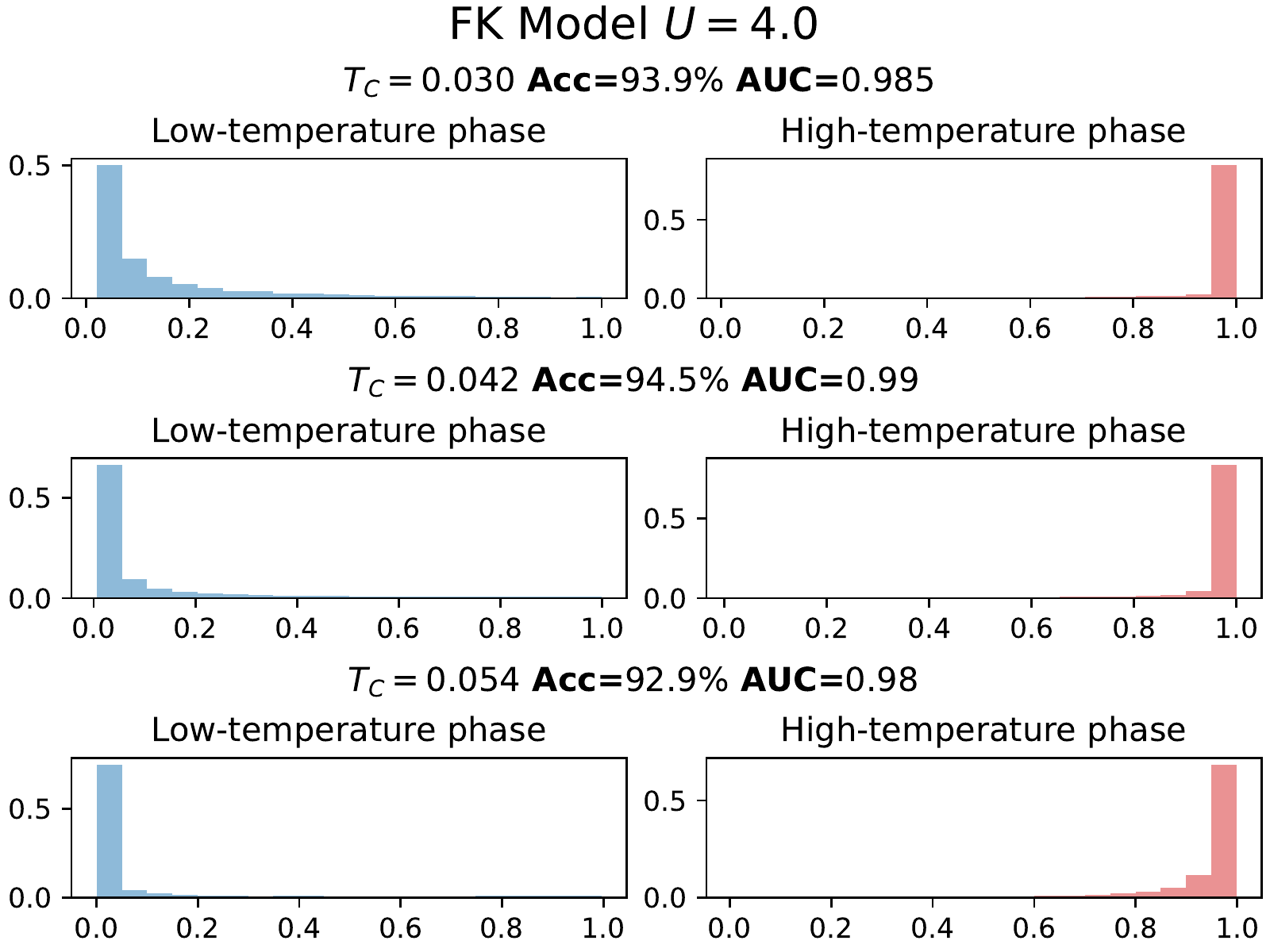}
    \caption{Distributions of outputs of the sigmoid activation function $\sigma(x)$ obtained from the neural network fed with the spin configurations generated for FK model $U=4.0$ (continuous phase transition) in the case of different critical temperatures $T'_{c}$}
    \label{fig:U=4.0_auc}
\end{figure}

As described in the main text, the evaluation of the LbC scheme requires the calculation of the performance of a neural network at different critical temperatures $T'_{c}$. In most cases, the input to this neural network is imbalanced, i.e. the number of samples assigned to one category is larger than to the other one. 
Such an imbalanced dataset is not a problem when we aim at finding the true critical temperature -- the LbC scheme was originally built on that assumption. However, when it comes to the more subtle problem of determining the character of a phase transition, our predictions should be more accurate. In Ref. \cite{huang} it has been proven, based on objective criteria of statistical consistency and discriminancy, that AUC-ROC metrics is generally better metrics in this case. In addition, we are convinced that, in the context of our research, it is important to gain information not only about the accuracy of predictions, but also about the distribution of the neural network's output. For instance, if we measure the performance of a model with accuracy, for a sample belonging to a class labeled as $1$, it does not make any difference if the output of a model is $0.55$ or $0.99$ - as long as it is above threshold value $0.5$ it is considered as a correct prediction. On the other hand, if we choose AUC--ROC metrics instead, we obtain information not only about how good a model is at classification, but also about the degree of separability of the data assigned to different categories. This fact is illustrated in Figs.~\ref{fig:U=4.0_auc} and ~\ref{fig:U=0.5_auc} which present the distributions of results obtained from the output of a neural network in the case of a model exhibiting the first--order phase transition (Fig.~\ref{fig:U=0.5_auc}) and second--order phase transition (Fig.~\ref{fig:U=4.0_auc}) at various critical temperatures $T'_{c} < T_{c}$, $T'_{c}\approx T_{c}$, and $T'_{c} > T_{c}$, respectively. It is clearly seen that although the temperature difference $\Delta T = 0.012$ between consecutive panels is the same in both cases, the distribution of the results is quite different. When it comes to the first--order phase transition, the outputs are almost perfectly separated for the alleged critical temperature $T'_{c} < T_{c}$ and $T'_{c}\approx T_{c}$. However, this perfect separability is not observed for temperature $T'_{c} > T_{c}$, where the outputs for the high-temperature phase are more ``smeared out''. The situation is different when it comes to the second--order phase transition (\ref{fig:U=4.0_auc}). In such a case, the distribution of the results is basically the same in all three occurrences (although  the separation is slightly more pronounced for $T'_{c}\approx  T_{c}$).

\section{Defects\label{defect_defs}}
\begin{figure}[h]
    \centering
    \vskip 2mm
    \includegraphics[width=0.85\columnwidth]{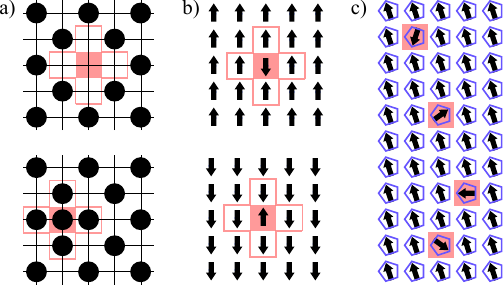}
    \caption{Definitions of defects (marked in red) for the FK model (a), the BC model (b), and the 5-state Potts model (c). Note, that for the Potts model the defects shown correspond only to one particular direction of the global magnetization.}
    \label{fig:defect_defs}
\end{figure}

Figure \ref{fig:defect_defs} illustrates the definitions of defects in otherwise perfectly ordered states for all the models studied here. By a ``defect'' we understand any lattice site that does not match its neighbors in the ordered phase. In the case of the FK and BC models, there is only one possible kind of defect (empty/occupied site for the FK model and spin up/spin down site for the BC model), while for the {\textit q}P model for any direction of global magnetization $s$ ($s=1,\ldots,q$) there are $q-1$ possible defects with spin $s'$ such that $s'\ne s$.

\bibliography{apssamp}

\end{document}